\documentclass[fleqn,usenatbib]{mnras}
\usepackage{hyperref}
\usepackage{multirow}
\usepackage{graphicx}
\usepackage{times}
\usepackage{float}
\usepackage{pdflscape}
\usepackage{threeparttable}
\usepackage{bm}
\usepackage{amsmath}
\usepackage{amssymb}
\usepackage{xspace}
\usepackage{newtxtext}
\usepackage{color}
\usepackage{tikz}
\usetikzlibrary{positioning}

\usepackage{graphicx}
\usepackage{txfonts}
\usepackage{gensymb}
\usepackage[]{hyperref}
\usepackage{multirow}
\usepackage{graphicx}
\usepackage{times}
\usepackage{float}
\usepackage{pdflscape}
\usepackage{threeparttable}
\usepackage{amsmath}
\usepackage{amssymb}
\usepackage{xspace}
\usepackage{caption}
\usepackage{subcaption}
\usepackage{newtxtext}
\usepackage{color}

		\newcommand{\Msun}{\mbox{$M_{\odot}$}\xspace}

		\newcommand{\feka}{Fe\,K$\alpha$\xspace}		
		 
		\newcommand{\fexxvxxvi}{Fe\,\textsc{xxv--xxvi}\xspace}

		\newcommand{\fexxvabs}{Fe\,\textsc{xxv}\,He$\alpha$\xspace}
		\newcommand{\fexxvi}{Fe\,\textsc{xxvi}\xspace}
		\newcommand{\fexxviabs}{Fe\textsc{xxvi}\,Ly$\alpha$\xspace}

		\newcommand{\dchidof}{\Delta\chi^{2}/\Delta\nu}
		\newcommand{\chidof}{\chi^{2}/\nu}
		
		\newcommand{\nhgal}{N_{\rm H}^{\rm Gal}}
		\newcommand{\nh}{N_{\rm H}}
		\newcommand{\lognh}{\log(N_{\rm H}/\rm{cm}^{-2})}

		\newcommand{\lbol}{L_{\rm bol}} 
		\newcommand{\kbol}{k_{\rm bol}}

		\newcommand{\ledd}{L_{\rm Edd}}
		\newcommand{\eddr}{\lambda_{\rm Edd}}
		\newcommand{\medd}{\dot M_{\rm Edd}}
		\newcommand{\mbh}{M_{\rm BH}}
		\newcommand{\pout}{\dot p_{\rm out}}
		
		\newcommand{\prad}{\dot p_{\rm rad}}

		\newcommand{\rout}{R_{\rm out}}
		\newcommand{\rmax}{R_{\rm max}}
		\newcommand{\rmin}{R_{\rm min}}
		
		\newcommand{\fig}{Fig.\,}

		\newcommand{\ks}{\,\rm ks}
		
		\newcommand{\dchis}{\Delta\chi^{2}/\Delta\nu}

		\newcommand{\fv}{f_{\rm v}}
		\newcommand{\lx}{\mathcal{L}_{\rm X}}
		\newcommand{\los}{line-of-sight\xspace}
		\newcommand{\vw}{v_{\rm w}}

		\newcommand{\vout}{v_{\rm out}\xspace}
		\newcommand{\mout}{\dot M_{\rm out}}
		\newcommand{\mw}{\mathcal{\dot M}_{\rm w}}			
		\newcommand{\vesc}{v_{\rm esc}}
		\newcommand{\vinf}{v_{\infty}}

		\newcommand{\mcrt}{$\mathcal{MCRT}$\xspace}
		\newcommand{\bxa}{\textsc{bxa}\xspace}
		\newcommand{\ann}{\texttt{ANN}\xspace}

		\newcommand{\fe}{Fe\,K\xspace}
        \newcommand{\iron}{iron\,K\xspace}
		\newcommand{\ergs}{\,\rm erg\,s^{-1}\xspace}
		\newcommand{\cmsq}{{\,\rm cm^{-2}}\xspace}

		\newcommand{\rg}{\,r_{\rm g}}
		\newcommand{\ev}{\,\rm eV}		
		\newcommand{\kev}{\,\rm keV}

		\newcommand{\ew}{\rm EW\xspace}

		\newcommand{\athena}{\emph{Athena}\xspace}
		\newcommand{\xrism}{\emph{XRISM}\xspace}
		\newcommand{\athenaifu}{\emph{Athena}/X-IFU\xspace}
		\newcommand{\xrismres}{\emph{XRISM}/Resolve\xspace}
		 
		\newcommand{\nustar}{\emph{NuSTAR}\xspace}		
		\newcommand{\xmm}{\emph{XMM-Newton}\xspace}
		
		\newcommand{\xmmrgs}{\emph{XMM-Newton} RGS\xspace}

		\newcommand{\xmmnu}{\textit{XMM-Newton} and \textit{NuSTAR}\xspace}

		\newcommand{\pds}{PDS\,456\xspace}		
		\newcommand{\mcg}{MCG--03--58--007\xspace}

		\newcommand{\xrade}{\textsc{xrade}\xspace}

		\newcommand{\optxagn}{\texttt{optxagnf}\xspace}

		\newcommand{\dws}{\texttt{slow64}\xspace}
		\newcommand{\dwf}{\texttt{fast32}\xspace}
		
		\newcommand{\xstar}{\textsc{xstar}\xspace}
		\newcommand{\xspec}{\textsc{xspec}\xspace}

\tikzset{%
  every neuron/.style={
    circle,
    draw,
    minimum size=0.8cm
  },
  neuron missing/.style={
    draw=none, 
    scale=2,
    text height=0.3cm,
    execute at begin node=\color{black}$\vdots$
  },
}

\title[\textit{XRADE}]{A New Emulated Monte Carlo Radiative Transfer Disk-Wind Model:\\ 
X-Ray Accretion Disk-wind Emulator -- \textsc{XRADE}}

\author[Matzeu et al.]{G. A. Matzeu$^{1,2,3}$\thanks{Correspondence to: gabriele.matzeu@unibo.it},
M. Lieu$^{4}$, M. T. Costa$^{5}$, J. N. Reeves$^{6,7}$, V. Braito$^{7,6}$, M. Dadina$^{2}$, 
\newauthor
E. Nardini$^{8}$,
P. G. Boorman$^{9,10}$,
M. L. Parker$^{11}$, 
S. A. Sim$^{12}$,  
D. Barret$^{13}$,
E. Kammoun$^{13,8}$,
\newauthor
R. Middei$^{14}$,
M. Giustini$^{15}$,
M. Brusa$^{1,2}$,  
J. Pérez Cabrera$^{16}$, 
S. Marchesi$^{2,17}$ 
\\
$^{1}$Department of Physics and Astronomy (DIFA), University of Bologna, Via Gobetti, 93/2, I-40129 Bologna, Italy\\
$^{2}$INAF-Osservatorio di Astrofisica e Scienza dello Spazio di Bologna, Via Gobetti, 93/3, I-40129 Bologna, Italy\\
$^{3}$European Space Agency (ESA), European Space Astronomy Centre (ESAC), E-28691 Villanueva de la Ca\~{n}ada, Madrid, Spain\\
$^{4}$ School of Physics $\&$ Astronomy, University of Nottingham, Nottingham, NG7 2RD\\
$^5$Astrophysics Group, School of Physical and Geographical Sciences, Keele University, Keele, Staffordshire ST5 5BG, UK\\
$^{6}$Department of Physics, Institute for Astrophysics and Computational Sciences, The Catholic University of America, Washington, DC\,20064, USA \\
$^{7}$INAF -- Osservatorio Astronomico di Brera, Via Bianchi 46, I-23807 Merate (LC), Italy\\
$^{8}$INAF -- Osservatorio Astrofisico di Arcetri, Largo Enrico Fermi 5, I-50125 Firenze, Italy\\
$^{9}$Astronomical Institute of the Czech Academy of Sciences, Bo\v cn\'i II 1401/1A, 14100 Praha 4, Czech Republic\\ 
$^{10}$School of Physics \& Astronomy, University of Southampton, Highfield, Southampton SO17\,1BJ, UK\\ 
$^{11}$Institute of Astronomy, University of Cambridge, Madingley Road, Cambridge, CB3 0HA, UK\\
$^{12}$School of Mathematics and Physics, Queen's University Belfast, Belfast, UK\\
$^{13}$IRAP, Universit\'e de Toulouse, CNRS, UPS, CNES 9, Avenue du Colonel Roche, BP 44346, F-31028, Toulouse Cedex 4, France\\
$^{14}$Space Science Data Center - ASI, Via del Politecnico s.n.c., 00133 Roma, Italy\\
$^{15}$Centro de Astrobiolog\'ia (CSIC-INTA), Camino Bajo del Castillo s/n,  
Villanueva de la Ca\~{n}ada, E-28692 Madrid, Spain\\
$^{16}$Aurora Technology B.V., European Space Astronomy Centre (ESAC),  Camino Bajo del Castillo s/n, Villanueva de la Ca\~{n}ada, E-28692 Madrid, Spain\\
$^{17}$Department of Physics and Astronomy, Clemson University, Kinard Lab of Physics, Clemson, SC 29634, USA}

\begin{document}

\date{\today}

\pagerange{\pageref{firstpage}--\pageref{}} \pubyear{?}

\maketitle
\label{firstpage}

\maketitle
\begin{abstract}

We present a new X-Ray Accretion Disk-wind Emulator (\xrade) based on the 2.5D Monte Carlo radiative transfer code which provides a physically-motivated, self-consistent treatment of both absorption and emission from a disk-wind by computing the local ionization state and velocity field within the flow. \xrade is then implemented through a process that combines X-ray tracing with supervised machine learning. We develop a novel emulation method consisting in training, validating, and testing the simulated disk-wind spectra into a purposely built artificial neural network. The trained emulator can generate a single synthetic spectrum for a particular parameter set in a fraction of a second, in contrast to the few hours required by a standard Monte Carlo radiative transfer pipeline. The emulator does not suffer from interpolation issues with multi-dimensional spaces that are typically faced by traditional X-ray fitting packages such as \xspec. \xrade will be suitable to a wide number of sources across the black-hole mass, ionizing luminosity, and accretion rate scales. As an example, we demonstrate the applicability of \xrade to the physical interpretation of the X-ray spectra of the bright quasar \pds, which hosts the best-established accretion-disk wind observed to date. We anticipate that our emulation method will be an indispensable tool for the development of high-resolution theoretical models, with the necessary flexibility to be optimized for the next generation micro-calorimeters on board future missions, like \xrismres and \athenaifu. This tool can also be implemented across a wide variety of X-ray spectral models and beyond. 
\end{abstract}


\section{Introduction}

Accretion-disk winds are generally observed through blueshifted absorption features at rest-frame energies $>7\kev$, imprinted in the X-ray spectra of active galactic nuclei \citep[AGNs;][]{Chartas02,Reeves03,Pounds03}. Their degree of blueshift from the lab energies of \fexxvabs (He-like) and/or \fexxviabs (H-like) implies mildly relativistic outflow velocities, typically falling in the range $\sim0.1$--$0.4c$ \citep[e.g.,][]{Reeves09,Gofford14,Nardini15,Matzeu16,Matzeu19,Parker20iras,Middei20}. Their frequent detection, in approximately $35$--$40\%$ of local AGNs \citep{Tombesi10,Gofford13,Igo20}, suggests that the wind geometry is characterized by a large covering factor ($\Omega$). This was confirmed by the direct measurement of $\Omega\gtrsim2\pi$ in the luminous quasar \pds \citep[][N15 hereafter]{Nardini15}. With such a high covering factor, coupled with high column densities ($\nh\gtrsim10^{23}\cmsq$, \citealt{Tombesi11, Gofford13}) and high velocities, a large amount of kinetic power can be transported, possibly exceeding the $0.5$--$5\%$ of the bolometric luminosity required for significant AGN feedback \citep{King03,KingP03,DiMatteo05,HopkinsElvis10}. 

Measuring the intrinsic physical properties of these winds can provide important insights into the mechanism through which they are driven (launched, accelerated). There are currently three known physical mechanism responsible for driving accretion disk winds: gas pressure, radiation pressure, and magnetic fields. While gas pressure (thermal driving) is unable to explain the large velocities observed in accretion disk winds in AGN, the two other mechanisms are in principle able to do so.
Three possible scenarios might therefore be able to explain the observations of AGN accretion disk winds: (i) \textit{radiatively-driven} winds \citep[e.g.,][]{Proga00,Proga04,Kallman01,Giustini19}; (ii) \textit{magnetically-driven} (MHD hereafter) winds \citep[e.g.,][]{1992ApJ...385..460E,Ohsuga09,Fukumura10,Kazanas12,Fukumura15}; and/or (iii) to some extent a likely combination of the two \citep[e.g.,][]{1995ApJ...455..448D,2005ApJ...631..689E,Matzeu16}.

In the radiatively driven scenario, the AGN radiation pressure launches a wind from the accretion disk from tens to thousands of gravitational radii from the supermassive black hole (SMBH; the gravitational radius $r_g = G\mbh/c^{2}$, with $G$ the gravitational constant, $c$ the speed of light, and $\mbh$ the black hole mass). The detection of strongly blueshifted broad absorption lines (BALs), associated with the UV transitions \citep[e.g.,][]{1991ApJ...373...23W, Matthews16,2020MNRAS.492.4553R} demonstrates that substantial momentum can be transferred from a powerful radiation field to the gas, thus accelerating mass outflows. These type of radiatively driven outflows are described as \textit{line-driven} winds, as their strength depends on the opacity of the absorption lines, which acts as a \textit{force multiplier} to the radiation pressure and can make the bound-bound absorption cross-section considerably larger than the Thomson cross-section for electron scattering \citep[i.e., $\sigma_{line} \gg \sigma_T$; e.g.,][]{Castor75,Stevens90,Dannen19}.
The strength of line-driven disk winds depends  on the ionization state of the gas $\xi=L/nR^2$, where $n$ is the gas density, $L$ is the ionising luminosity, and $R$ is the distance between the gas and the source of the ionising luminosity. As demonstrated by \citet{Dannen19}, for a typical AGN spectral energy distribution the effects of the force multiplier drop at $\log\xi > 3$, where all the relevant opacity is lost. Line-driven winds are therefore likely more relevant for sub-Eddington sources\footnote{The Eddington Luminosity is defined as  $L_{Edd}=4\pi\, GMm_p c/\sigma_T$, with $m_p$ the proton mass and $\sigma_T$ the Thomson cross section, and it is the luminosity for which  the radiation pressure and the gravitational pull are equal, for a given mass $M$.}, where the ionising luminosity is not as large as completely ionise the illuminated gas.

In AGN close to Eddington or super-Eddington, the ionization state of the gas is so high that the dominant interaction between the outflowing gas and the radiation field is likely Thomson (and Compton) scattering \citep{KingP03,King10}. In this case, a direct correlation between the momentum rate of the outflow and the momentum rate of the radiation field, i.e.
$\pout\sim\prad\,(=L/c)$, would be expected if the optical depth to electron scattering is $\tau\sim1$.
This indeed appears to be the case in many observations of fast, highly-ionized winds \citep{Tombesi13,Gofford15,Nardini19}, but it does require the AGN to radiate at a considerable fraction of its Eddington luminosity, $\ledd$ \citep{KingP03}. 

Most theoretical outflow studies are mainly concentrated on radiatively-driven winds in both AGNs \citep{Sim08,Sim10,Hagino15,Hagino16b,Hagino16a,Nomura16,Matthews16,Matthews20,Luminari18,Nomura20,QueraBofarull20,Mizumoto21}, X-ray binaries \citep[XRBs;][]{Higginbottom19,Higginbottom20,Tomaru20,Tomaru20b}, and cataclysmic variables \citep[e.g.,][]{Matthews15}. Nevertheless, MHD wind models have been successfully applied to both AGNs \citep{Fukumura10,Fukumura15,Fukumura18} and XRBs \citep{Fukumura17,Fukumura21,Ratheesh21}. These findings suggest that both driving/launching mechanisms apply across the black hole mass and luminosity scales. 


The development of physical models for accretion disk winds and a self-consistent test of their predictions are among the primary goals in modern X-ray astronomy.
Predictions can be tested by using grids of spectral simulations generated for different values of the physical parameters of interest, such as the ionisation state and column density of the gas. Up until now, astronomers had to compromise between the sampling resolution and the extent of the parameter space covered in the model, due to the extremely demanding computational times involved. Although grids generated with coarser sampling generally allow one to explore a broader parameter space, they are more susceptible to interpolation issues \citep{Arnaud96} that may affect the degree of accuracy of the measurements.
The next generation of instruments, on board \textit{XRISM} and \athena (planned to be launched in 2023 and early 2030's, respectively) will provide a significant increase in spectral resolution, with $\Delta E\sim5\ev$ for \xrismres \citep{Tashiro20XRISM} and $\Delta E\sim2.5\ev$ for \athenaifu \citep{Barret18}. Such advances in technology will inevitably require the development of higher resolution grids to match the improved spectral information.

Machine learning techniques, which allow us to learn the mapping from an input space to an output space, can play a fundamental role in speeding up this process. In supervised machine learning, a sample of both the input and the output is known, and the objective is to learn a mapping that is able to best reproduce the output for a given {\it loss function}. In our case, the loss function is a measure of how close the machine learning emulated X-ray spectra are to {\it ground truth}, the simulated spectral values. This is the training of the model, and the data sample is known as the training data. Machine learning benefits from large data samples and although such process can be computationally expensive to train, the trained data are capable of efficiently computing the mapping. 

Consequently, supervised learning methods can be useful to reduce the computational cost of large and complex models, provided that a representative training set can be obtained \citep{kasim2020}. Trained machine learning models can be used as surrogate models to approximate computationally expensive models such as the weather and climate \citep{watson2021}. These frameworks are known as {\it emulators}, and they can be developed as \textit{artificial neural networks} (\texttt{ANN}s hereafter). The \ann architecture is loosely based on the human brain and consists of interconnected neurons organized into layers. \texttt{ANN}s are also quickly becoming popular in astronomy to approximate simulations and interpolate between them \citep[see e.g.][]{chardin2019, he2019}. \citet{kerzendorf21} created an emulator to replace expensive radiative transfer codes for modelling supernova spectral time series, while \citet{Alsing2020}'s stellar population synthesis (SPS) model emulator accurately generates galaxy spectra and photometry from SPS parameters. 

The aim of this paper is to present the description, development, and application of a new extended Monte Carlo radiative transfer (\mcrt hereafter) accretion disk-wind code initially developed by \citet[][S08, S10 hereafter]{Sim08,Sim10}: X-Ray Accretion Disk-wind Emulator (\textsc{xrade}). The novel approach in the development of \xrade is twofold: (i) firstly, we compute a new set of synthetic X-ray spectra in order to explore the physical conditions of accretion disk-winds in a larger AGN population. (ii) Secondly, as the synthetic spectra are fed into a purposely built \ann, the data will undergo a process of training, validation, and testing with the aim of: (a) accelerating the process of synthetic spectra simulations, and (b) solving the multi-dimensional interpolation problems\footnote{\url{https://heasarc.gsfc.nasa.gov/xstar/docs/html/node95.html}} that arise when multiplicative tables are adopted in spectral fitting packages such as \xspec \citep{Arnaud96}. Our \ann allows the user to generate customised \xrade tables at their requirement. On this basis, \textbf{we generated two new large \mcrt tables,} namely \dws and \dwf, which cover a larger parameter space (CPU time: $\sim$7--8 months with 480 50\,Gb cores), than the one generated in S08, S10 and \citet{Reeves14}. In the future, the spectral resolution of the wind grids will also be increased, in order to match the next-generation calorimeter data. Hence to reduce the computational demands for our future tables, machine learning is a very important tool. 

This paper is organized as follows: in \autoref{sec:model overview} we give an overview of the \mcrt methods used to simulate disk-wind synthetic spectra from the code originally developed in S08, S10, and we describe the physical assumptions adopted in the disk-wind \dws and \dwf models. We also discuss the input parameters and we present a brief description of the main input parameters. In \autoref{sec:neural emulator} we describe in detail the methods adopted for the development of \xrade. In \autoref{sec:application on pds456} we apply \xrade to the quasar \pds, which hosts one of the most powerful and persistent accretion disk-winds discovered to date. We specifically test \xrade on the \xmmnu 2013 observation of \pds, as the X-ray spectrum is characterized by the most prominent and best studied P-Cygni-like profile 
yet observed. In \autoref{sec:conclusion and future work} we draw our conclusions and discuss further work.

\begin{table*}
\caption{\dwf and \dws input parameters. Note that the $172,800$ output spectra will be available for developing \xrade. For the quantities flagged with ${\dagger}$ or ${\ddagger}$, more details are provided in Appendix\,\ref{sub:velocity_subsection} and  \ref{sub:Mass density} respectively. Note that in the bottom 5 rows the \ann input parameters are also the measurable output when \dwf and \dws tables are loaded into X-ray fitting packages such as \xspec.}
  \begin{tabular}{ l | c r }


\hline
             Input Parameter  &       \multicolumn{2}{c}{Values}  \\

                            & \dwf & \dws  \\
\\

    
range of source photon energies in simulation                   &\multicolumn{2}{c}{$0.1$--$511\kev$} \\

photon packets                                                  & \multicolumn{2}{c}{$160,000$} \\

size of X-ray emission region\,($r_{\rm er}$)				    &\multicolumn{2}{c}{$6\rg$}\\
	
inner radius of the accretion disk\,($r_{\rm d}$)               &\multicolumn{2}{c}{$6\rg$}\\

inner launch radius\,($\rmin$)								    &$32\rg$                               &$64\rg$  \\

outer launch radius\,($\rmax$)                                  &\multicolumn{2}{c}{$1.5\rmin$}  \\

distance to wind focus\,($d$) 								    &\multicolumn{2}{c}{$\rmin$} \\

velocity scale length\,$(R_{v})^{\dagger}$						&\multicolumn{2}{c}{$\rmin$} \\

velocity exponent\,$(\beta)^{\dagger}$							&\multicolumn{2}{c}{$1.0$} \\

launch velocity\,$(v_{0})^{\dagger}$							&\multicolumn{2}{c}{$0.0$} \\

mass-loss exponent\,$(\kappa)^{\ddagger}$						&\multicolumn{2}{c}{ $-1.0$}\\

outer radius of simulation grid 							    &\multicolumn{2}{c}{$33876\rg$}\\

3D Cartesian RT grid cells 									    &\multicolumn{2}{c}{$180\times180\times180$}\\

2D wind grid zones 											    &\multicolumn{2}{c}{$100\times100$}\\

\hline

Input Parameter  &       \multicolumn{2}{c}{Values}  \\
\\
source power-law photon index\,($\Gamma$)                        &\multicolumn{2}{c}{$\{1.6, 1.8, 2.0, 2.2, 2.4 \}$} \\
terminal velocity parameter\,($\fv$) 							   &\multicolumn{2}{c}{$\{0.25, 0.50, 0.75, 1.0, 1.25, 1.50, 1.75, 2.0 \}$}  \\
source luminosity\,$\left(\lx=\frac{L_{2-10\kev}}{\ledd}\right )$            &\multicolumn{2}{c}{$\{0.0252, 0.0475, 0.0796, 0.1415, 0.2516, 0.4475, 0.7958, 1.4151, 2.5165 \}\,\times \,10^{-2}$} \\
wind mass-loss rate\,$\left( \mw=\frac{\mout}{\medd}\right )$			       
&\multicolumn{2}{c}{$\{0.0196, 0.0270, 0.0373, 0.0515, 0.0710, 0.0980, 0.1352, 0.1866, 0.2575, 0.3552, 0.4901, 0.6762\}$ } \\
angular bins\,($\mu=\cos\theta$)			       
&\multicolumn{2}{c}{$\{0.025, 0.075, 0.125, 0.175, 0.225, 0.275, \cdots  0.725, 0.775, 0.825, 0.875, 0.925, 0.975   \}$ } \\

  \hline
  \end{tabular}
\label{tab:in_parameters}
\end{table*}

\section{Radiative Transfer Code overview}
\label{sec:model overview}

The development of \xrade is based on training the 
synthetic wind spectra simulated with the \mcrt code 
by S08, S10 into an \ann. 
Note that a more detailed description of the input model setup can be found in S08, S10. In this section we present, for completeness, an overview of the physical basis and approach adopted in generating the input \mcrt wind spectra for \xrade.

\begin{figure}
  \includegraphics[width=\linewidth]{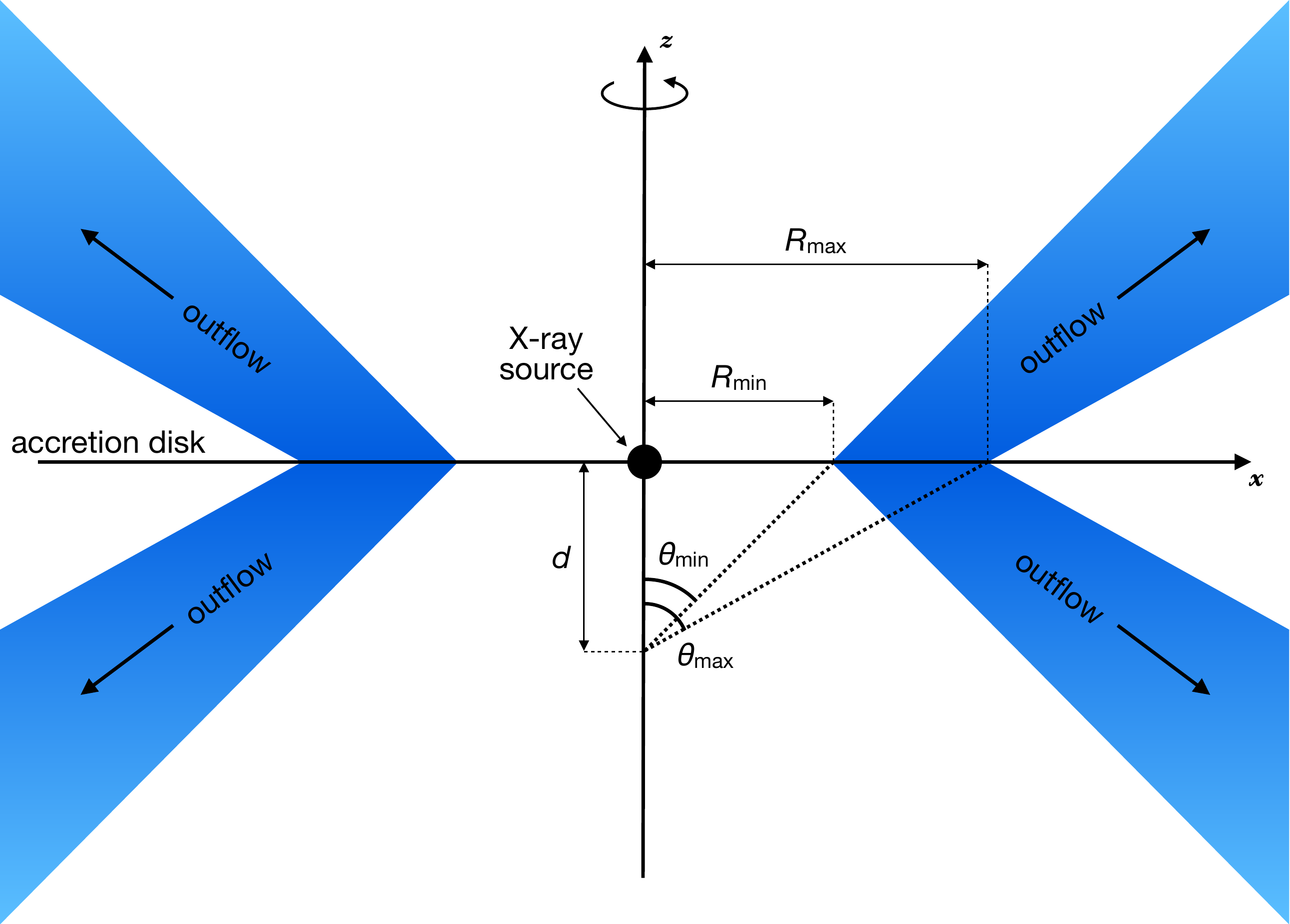}
  \caption{ Schematic representation of the bi-conical structure adopted for the disk-wind model. The 2D geometry is defined with three parameters: $\rmin$, 
  $\rmax$, and $d$. $\rmin$ and $\rmax$ correspond to the radii at which the inner and outer edges of the disk wind intercept the disk plane, respectively. The blue-shaded area represents the physical extent of the outflow, while $d$ defines the focus point of the wind below the disk plane, which controls the degree of collimation and opening angle of the wind (where $\tan \theta_{\rm min}=\frac{\rmin}{d}$). This 2D structure is then rotated around the vertical $z$--axis and mirrored with respect to the accretion-disk plane to produce an axisymmetric 2.5D wind geometry.}
  \label{fig:Diskwind_KWDSchema}
\end{figure}

Initially, S08 carried out multi-dimensional (2.5D) Monte Carlo radiative transfer simulations in a bi-conical wind structure (see \autoref{fig:Diskwind_KWDSchema}). The simulated spectra were calculated over grid points with coordinates $x,\,y,\,z$, under the assumption that the system is axisymmetric about the polar $(z)$ axis in the azimuthal direction. S10 extended the atomic database with the inclusion of the L- and M- shell transitions. As a result, the simulated synthetic spectra were more accurate over a larger range of photon energies (i.e., $0.2$--$10\kev$). Additionally, the Monte Carlo ray-tracing method described in \citet{Lucy02,Lucy03} was implemented in the code. This allowed the treatment of ionization and radiative heating of the gas by means of self-consistent calculations of the heating/cooling of electrons based on the \textit{photon packets} (the computational structure used in the simulations) that propagate throughout the wind. A temperature gradient is then calculated in order to provide a more physical representation of the ionization structure of the wind. This process is then reiterated multiple times to accurately define the heating and cooling rates for the wind until they reach equilibrium. In \dws and \dwf we set $160,000$ photon packets each, which are then collected and grouped into $10,000$ energy bins (and binned up by a factor of 10$\times$ giving a total of 1000 energy bins), based on their viewing (observer) angle ($\theta$) from the $z$-axis. 

The \mcrt code creates tables of simulated wind spectra that take into account the effects of the radiation transmitted through the wind, which include the scattering and reflected emission from the flow. An interesting outcome of this model is that the accretion disk-wind itself can give rise to \fe emissions, with line widths up to $\sigma_{\rm width}\sim1$\,keV \citep{Parker22arxivdiskwind} . Such profiles are obtained from the combination of: (i) velocity shear in the flow, (ii) its rotation around the polar axis, and (iii) the Compton scattering of the Fe\,K$\alpha$ line photons in the wind. Thus the disk-wind model can provide a physically-motivated, self-consistent treatment of both the absorption and emission produced in the wind by computing the ionization state and velocity field within the flow. In other words, the disk-wind model calculates the ionization at each point in the wind over a wide range of states, thus describing a more realistic (non-uniform) ionization structure and velocity ﬁeld throughout the outflow. As for iron, the code covers charge states from Fe\,\textsc{x--xxvii}, and the output spectra include not just the absorption and emission from \fe, but also from L-shell iron and the K-shell lines of lighter elements in the soft X-ray band.

Routines that take into account special relativistic aberration of angles and Doppler shifts between the co-moving and observer frames are included in our \mcrt code, so that the model fully accounts for special relativistic effects. Such an implementation provides realistic and accurate estimates of the mass outflow rate and overall energetics, as the local radiative pressure might require non-negligible special relativistic corrections \citep[e.g.,][]{Luminari20,Luminari21}. The number of energy packets used in each simulation is chosen so that the Monte Carlo noise in the estimators is $<$\,3 per cent. This level of precision is sufficient given the quality of the observational data available at present (see \autoref{sec:application on pds456}).

\subsection{Geometry}
\label{sub:Geometry}

The assumed bi-conical structure of the inner disk-wind geometry is shown in \autoref{fig:Diskwind_KWDSchema}. The $x$--axis corresponds to the plane of the accretion disk, and the $z$-- axis to the polar (rotational) direction. The black hole is located at the origin and the X-ray source is located within $6\rg$ from it (see \autoref{sub:Launch Radius}). $\rmin$ and $\rmax$ are, respectively, the distances from the origin to the inner and outer edge of the wind at the interception with the equatorial ($xy$) plane. The radii $\rmin$ and $\rmax$ (expressed in gravitational units)
then enclose the disk-wind launch region, and set the collimation and the overall opening angle (equatorial or polar) together with the parameter $d$, which represents the distance of the focal point of the wind along the $z$--axis below the origin. The overall wind inclination angle $\theta$ is measured with respect to the $z$--axis, with the polar opening angle defined as $\theta_{\rm min}=\arctan{\rmin/d}$. Here we set $d/\rmin = 1$, 
so the wind opening angle is $45$ degrees from the pole. The observer's polar angle is included in the code through $\mu=\cos \theta$, where any line of sight with $\mu$\,$<$\,0.7 intercepts the wind. The terminal velocity\footnote{See Appendix\,\ref{sub:velocity_subsection} for a calculation of the velocity field through the streamline, up to the maximum terminal velocity, $\vinf$.} attained by the wind is $\vinf=\fv\vesc$, where $\vesc=(2/\rmin)^{1/2}c$, and the factor $\fv$ parameter allows the user to vary the terminal velocity for a given launch radius (see below). The lines that extend from $d$ and intercept the $xy$--plane in $\rmin$ and $\rmax$ produce the first quarter of the bi-conical wind, which is made axisymmetric under rotation in the azimuthal direction and reflected with respect to the disk plane (see \autoref{fig:Diskwind_KWDSchema}). The difference between the outer- and inner-most launch radii ($\Delta R=\rmax-\rmin$) of the flow off the disk plane defines the overall thickness of the wind streamline.


\subsection{Velocity}
\label{sub:Velocity}

Having set up the geometric framework in which the wind is simulated, we now describe the properties and key parameters of the synthetic spectra which will be subsequently fed into the \ann. Note that the emulation process itself will be described in more detail in \autoref{sec:neural emulator}. For the purpose of this work we generated two \mcrt disk-wind wind tables named \dwf and \dws (see \autoref{tab:in_parameters} for the summary of their parameter space). The former is tuned to the fastest disk-wind cases like \pds, where typically $\vout/c=0.25$--$0.35$ \citep[e.g.,][]{Matzeu17b}, with $\rmin=32\rg$; thus, for $\fv=1,~\vinf=-0.25c$. The latter is instead tuned to slower winds, e.g., \mcg \citep[e.g.,][]{Braito22} or PG\,1211$+$143 \citep[e.g.,][]{Pounds16}, with $\rmin=64\rg$; for $\fv=1,~\vinf=-0.177c$ (see \autoref{sub:Launch Radius}). Our input choice of $\rmin$ is related to the range of outflow velocities typically observed in AGNs, between $\vw\sim0.05$--$0.4c$ \citep[e.g.,][]{Tombesi10,Gofford13,Reeves18PDS,Igo20,Chartas21}. The terminal velocity parameter $\fv$ can be considered a fine-tuning factor of the outflow velocity, which allows the user to adjust $\vinf$ to match their observations. So $\vinf$ is regulated by changing the $\fv$ parameter, for a given launch radius. Note that for these \mcrt simulations a black hole mass of $\mbh=10^{9}\Msun$ is assumed. However as most of the units are normalized, e.g. radii to the gravitational radius, mass outflow rate and X-ray luminosity to the Eddington value (see below), the output table parameters are black hole mass invariant.

Note that these versions of these \mcrt tables are newly generated in this work and they will be made publicly available. Hence, the new range of parameters are tabulated in \autoref{tab:in_parameters}. The spectral properties of these grids, in particular in relation to the inclination and launch radius, are discussed further in Appendix\,\ref{secapp:Spectral properties of the disk-wind}. Both the \dws and \dwf tables were generated with $\fv$ ranging between 0.25--2 in steps of $\Delta\fv=0.25$. As a result, the following ranges of $\vinf$ are covered:
%
\begin{equation} 
\vinf/c=
\begin{cases}
  	
  	-0.500 \lesssim \vinf/c \lesssim -0.0625 &\quad \rmin/\rg=32,\\
  
    
  	-0.354 \lesssim \vinf/c \lesssim -0.0442 &\quad \rmin/\rg=64.\\
\end{cases}   
\label{eq:vinf_range_limits}
\end{equation}  

For simplicity, for both the \dwf and \dws tables, the geometric thickness of the outflow is set to be $\rmax/\rmin=1.5$, but in principle this could be variable. The outer boundary of the simulations is set as $\log\,(\rout/\rg)=4.53$ (i.e., $\sim34,000\rg$), whereas the X-ray source is set to originate from a central region of $6\rg$ in radius. Both the \dws and \dwf tables are generated with 5 grid points for the photon index ($\Gamma$; see \autoref{sec:input_spectrum}), 8 for the terminal velocity parameter ($\fv$), 12 for the normalized mass outflow rate ($\mw$; see \autoref{sub:Mass outflow rate}), 9 for the ionizing luminosity ($\lx$; see \autoref{sub:Ionizing X-ray luminosity}) and 20 angular bins ($\mu$). The combination of these parameters produces $5\times8\times12\times9\times20=86,400$ synthetic spectra in each table, for a total of $172,800$. Each spectrum is simulated over $1000$\footnote{Note that 1000 energy bins are adopted when simulating CCDs resolution spectra i.e., $\Delta E=60\ev$ at $6\kev$, over the $0.1$--$511\kev$ range. For future micro-calorimeter resolution we will increase the binning by at least one order of magnitude.} spectral points, uniform in log-space, and subsequently used in the emulation process described below.

\subsection{Mass outflow rate} 
\label{sub:Mass outflow rate}

The mass within the flow is determined by the normalized mass outflow rate parameter, which is expressed in Eddington units as $\mw=\mout/\medd$ (a radiative efficiency for a Schwarzschild black hole of $\eta=0.06$ is assumed \citealt{ShapiroTeukolsky83book}). Hence, $\mw$ is not directly dependent upon the black hole mass of the source. An increase in $\mw$ affects the mass density in each cell by increasing the opacity of the medium thereby yielding a higher column density through the wind and deeper absorption lines (see Appendix\,\ref{sub:Mass density}). Additionally, as scattering of photons increases with opacity, the relative strength of the component scattered out of the flow would also increase proportionally with $\mw$. In both tables, the $\mw$ parameter covers the $0.020 < \mw < 0.676$ range in 12 equally-spaced logarithmic steps (see \autoref{tab:in_parameters}). This range covers the bulk of the typical measurements carried out in the literature i.e.,  $-2 \lesssim \log(\mw) \lesssim 0$ (see Fig. 2 in \citealt{Tombesi12} and Fig. 1 in \citealt{Gofford15}). Note that for future grids it is our intention to extend the $\mw$ parameter space to super-Eddington values, $\mw\gtrsim1$.

\subsection{Ionizing X-ray luminosity}
\label{sub:Ionizing X-ray luminosity}

The ionizing luminosity parameter is defined as the fraction of X-ray luminosity, calculated in the $2$--$10\kev$ band, with respect to the Eddington luminosity, i.e, $\lx=L_{2-10\kev}/\ledd$. As per $\mw$, with this normalization the $\lx$ parameter keeps the same meaning across the black-hole mass scale. $\lx$ measures the overall degree of ionization of the material within the flow, where lower values of $\lx$, typically $<1\%$ of $\ledd$, lead to the wind being less ionized and more opaque to X-rays. In contrast, an increase in $\lx$ will lead to winds that are more ionized and transparent to X-rays, to the extent that the spectrum becomes completely featureless. In the disk-wind code, the ionization of the plasma is self-consistently computed at each point in the wind, whilst both shielding and scattering of photons are also accounted for in the calculations. As a result, the overall ionization is stratified along the wind, whereby the innermost surface of the wind is almost fully ionized (mainly \fexxvi), as expected, being fully exposed to the X-ray source. The denser base of the wind is, not surprisingly, less ionized (with charge states down to Fe\,\textsc{x--xvi}). The decrease in ionization occurs both along the flow and across the base of the wind. More details regarding the input spectrum and its effect upon the wind ionization will be discussed in \autoref{sec:input_spectrum}. 

Compared to other models \citep[e.g.,][see \autoref{sub:other_modles}]{Hagino15}, the disk-wind code has access to more extensive atomic data, which cover a wide range in ionization; ions from Fe\,\textsc{x--xxvi} as well as from lighter elements such as C--Si are included. Thus, for any given observation of an AGN, the $\lx$ parameter can be calculated by comparing the intrinsic $2$--$10\kev$ luminosity to the (known) Eddington luminosity, and it is not a degenerate parameter in the modelling. The synthetic spectra for \xrade were simulated over a range of $2.5\times10^{-4} < \lx < 2.5\times10^{-2}$ (or $0.025\%$ to $2.5\%$ of $\ledd$) over 9 equally-spaced logarithmic increments (see \autoref{tab:in_parameters}). 
It is worth briefly discussing how such a range compares to the observed distributions of Eddington ratios ($\eddr=\lbol/\ledd$) and bolometric corrections ($\kbol=\lbol/L_{2-10\kev}$) as, by definition, $\lx=\eddr/\kbol$. These two quantities are known to correlate with each other, and their ratio typically falls in the range $\approx 10^{-3}-10^{-2}$ for the majority of type\,1 AGNs \citep[e.g.][]{Vasudevan09,Lusso12}. We conservatively adopt for $\lx$ a more extended range, especially at the low end, based on the evidence that the strongest winds are usually observed in sources that are relatively weak in the X-rays compared to the UV (hence a larger $\kbol$), which is interpreted as a requirement for effective line-driving \citep[e.g.,][]{Castor75,Giustini19}.


\subsection{The Input Spectrum}
\label{sec:input_spectrum}

The choice of the initial input spectrum is a crucial step for setting the \mcrt simulations, required for the development of \xrade, as the intrinsic spectrum can profoundly affect the observable disk-wind parameters. Steep (i.e., $\Gamma>2$) spectral slopes of the X-ray continuum are, in fact, critically responsible for producing strong absorption profiles. On the other hand, harder spectra (i.e., $\Gamma<2$) likely over-ionize the obscuring medium, leading to a considerable attenuation or disappearance of the absorption profiles \citep[e.g.,][]{Pinto18}. 

Various surveys on Seyfert galaxies and quasars  \citep[e.g.,][]{Porquet04b,Piconcelli05,Bianchi09caixa1,Scott14,Marchesi16,Williams18,Chartas21} established the diverse nature of the primary continuum slope in AGNs. The vast majority of objects studied in the above samples are type\,1 sources, hence they provide a reliable measure of their intrinsic spectral shape due to the general lack of obscuration. These studies show that $\sim$\,80\% of AGNs are characterized by an intrinsic slope distribution ranging between $\Gamma=1.6$--$2.4$ and peaking at $\Gamma\sim2$.

\begin{figure}
\centering
\includegraphics[width=\linewidth]{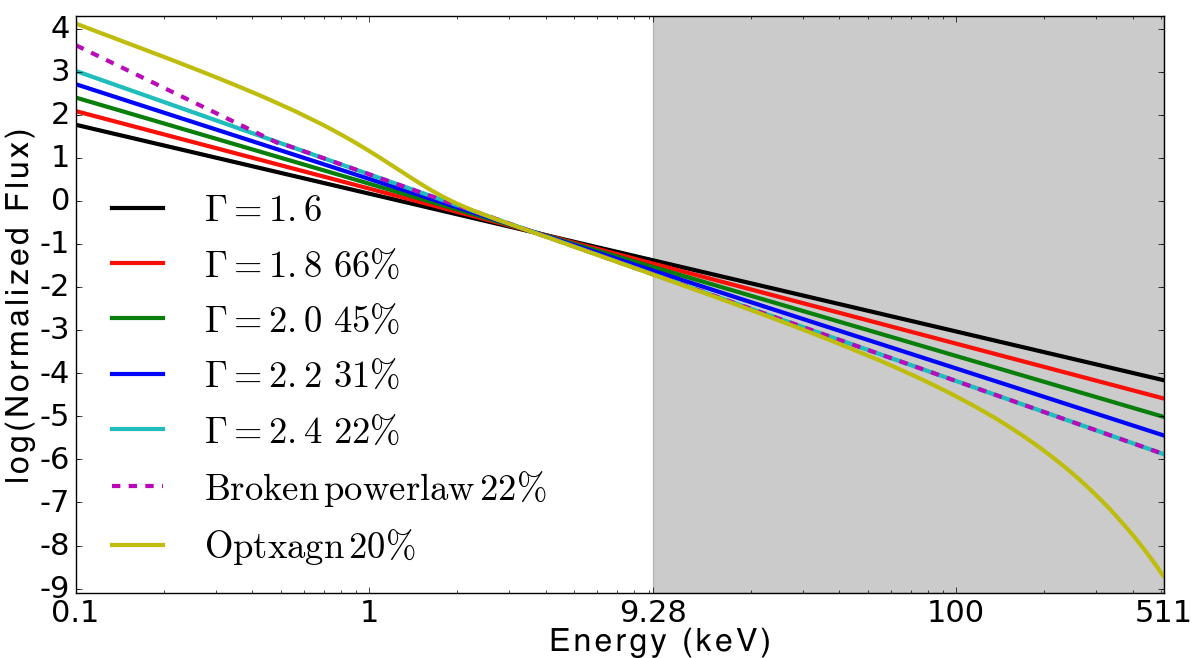}
\caption{Seven input SEDs, calculated in the $0.1$--$511\kev$ range, corresponding to five power laws with slope between $\Gamma=1.6$--$2.4$ and two more physically motivated input spectra, such as a double broken power law and the disk-corona Comptonization model \optxagn (see text). The spectra have been normalized to unity in the $2$--$10\kev$ band for comparison purposes. The shaded area indicates the energy band above the ionization threshold of \fexxvi, at $E=9.28\kev$.}
\label{fig:inputspec_renorm}
\end{figure}

A power-law SED is assumed to be a reasonable first-order approximation of the intrinsic X-ray continuum of AGNs, but in reality we know it to be much more complex. In \autoref{fig:inputspec_renorm} we show seven possible input spectra that correspond to five power-laws with $\Gamma=1.6$--$2.4$ along with two more complex SED models, such as a broken power-law and \optxagn\footnote{\optxagn is a self-consistent Comptonized disk emission model in \xspec, and it was adopted in generating \xstar \citep{BautistaKallman01,Kallman04} tables for \pds \citep[see Section\,4.2 in][for more details]{Matzeu16}. Note that in this exercise we adopted a $\Gamma=2.4$ and a hot coronal temperature of $kT_{\rm e}=100\kev$.} \citep[][]{Done12}, where the integrated $2$--$10\kev$ fluxes of the input spectra are normalized to unity. In this plot, the fraction of luminosity radiated above $E_{\rm lab}=9.28\kev$ (i.e., the ionization threshold of \fexxvi, shaded area) compared to the hardest ($\Gamma=1.6$) power law is calculated for each of the input continua. The percentages of the integrated photon flux in the $9.28$--$511\kev$ band, corresponding to each of the seven input spectra, are also noted in \autoref{fig:inputspec_renorm}. As the input spectrum becomes steeper, the number of photons above $E_{\rm lab}=9.28\kev$ decreases, leading to a lower mean charge of iron within the flow. On the other hand, harder spectra would induce a higher ionization of the gas, possibly over-ionizing iron for its K-shell to be significantly populated. 

\begin{figure}
\centering
\includegraphics[width=\linewidth]{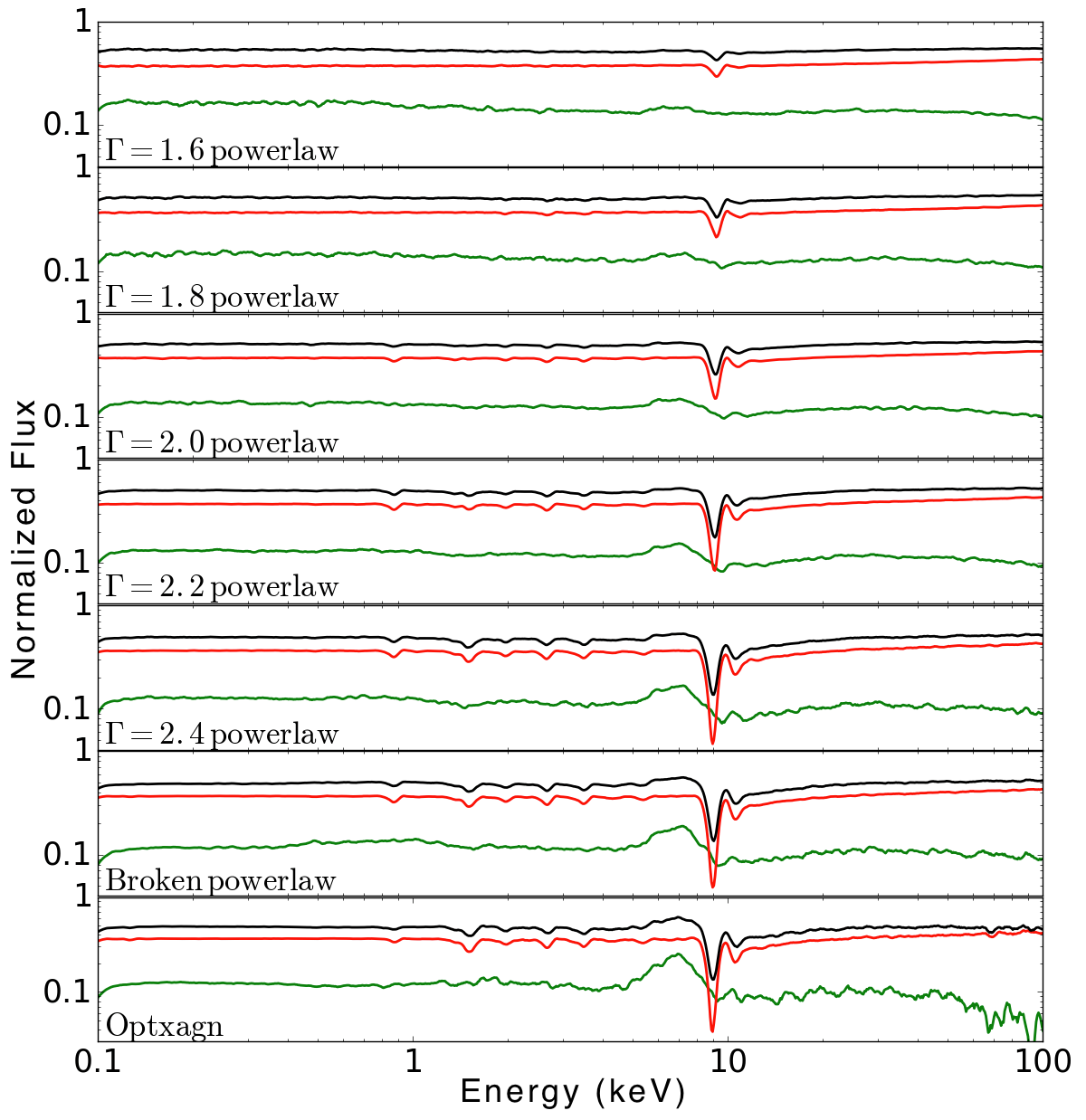}
\caption{The total (black), direct (red) and scattered (green) simulated wind spectra normalized to their corresponding input SED from \fig\ref{fig:inputspec_renorm}. The blueshifted absorption profile at $E\sim9\kev$ increases in strength and shifts to lower energies as the input SED becomes steeper. On the other hand, the hardening of the spectra would lead to an increase of ionizing photons that would eventually over-ionize the wind. As a consequence the over-ionized material would lose its opacity, which translates into a shallow absorption feature as shown in the top panel. For this example, the disk-wind simulations were carried out by assuming a $2$--$10\kev$ luminosity of 1.24\% of $\ledd$, an outflow rate of $\sim40\%$ of $\medd$, $f_v = 1.25\,(\vw=0.3125c~\rm for~\rmin=32\rg)$  and an inclination of $\mu=0.675$.}
\label{fig:change_sed}
\end{figure}

In \autoref{fig:change_sed}, we show the output spectra corresponding to the different $\Gamma=1.6$--$2.4$ in \autoref{fig:inputspec_renorm}, which illustrate how a change in ionization affects the spectra. Note that the \optxagn and broken power-law continuum, which both adopted a $\Gamma=2.4$ photon index at hard X-rays, produced a very similar \fe absorption line depth as per the corresponding simple power-law case. In other words, the cases with a more complex continuum (\optxagn, broken power-law) produced consistent results compared to the equivalent power-law case ($\Gamma=2.4$). Subsequently, to generate our \dws and \dwf tables, we choose a power-law SED with a photon-index range of $\Gamma=1.6$--$2.4$ with 5 linear steps of $\Delta\Gamma=0.2$ between $0.1$--$511\kev$. The above results in \autoref{fig:change_sed}, suggests that the strongest lines from disk-winds should occur in steep spectrum X-ray sources. For the case of the simulations in \autoref{fig:change_sed}, the equivalent width of the Fe\,\textsc{xxvi} line increases four-fold from $\Gamma=1.6$ ($\ew\sim110\ev$) to $\Gamma=2.4$ ($\ew\sim420\ev$). This could be the case observationally, where strong ($\ew\gtrsim100\ev$) blue-shifted \fe absorption lines are apparent in AGNs with steep ($\Gamma>2$) photon indices or when they are intrinsically X-ray weak (low $\lx$), e.g. \pds \citep{Reeves21}, PG\,1211$+$143 \citep{Pounds03}, IRAS\,13224$-$3809 \citep{Parker17,Pinto18}.

\begin{figure*}
\centering
\includegraphics[width=\linewidth]{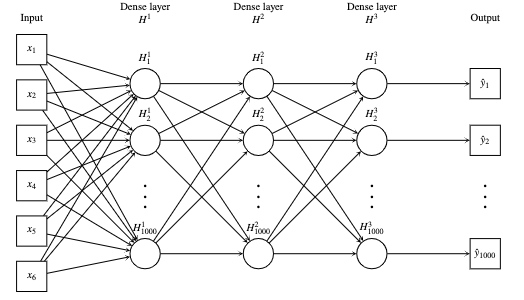}
\caption{The FFNN architecture of the emulator constructed for \xrade. Each circle represents a neuron. Each connection is marked as a right arrow.}
\label{fig:Architecture}
\end{figure*}

\section{Artificial Neural Emulator}
\label{sec:neural emulator}

\texttt{ANNs} are machine learning algorithms consisting of a set of neurons organized into layers. Each neuron is a distinct mathematical operation. They take an input $\mathbf{x}$ and apply an affine transformation followed by a threshold function $a$, known as the activation function, to ensure the mathematical operation is non-linear. This then allows several neurons to be applied sequentially, thus forming a network. If the network is fully connected then the output of each neuron in a given layer becomes the input to every neuron in the next layer: 
\begin{equation}\label{eqn:unit}
    x^{n}_{m} = a( \mathbf{x}^{n-1} \cdot \mathbf{W}_m + \mathbf{b}_m), 
\end{equation}
where $\mathbf{x}^{n-1}$ is the input to the $n$th layer and $x^n_m$ is the $m$th neuron in the $n$th layer. $\mathbf{W}$ and $\mathbf{b}$ are the trainable weight and bias (i.e., analogue role to a constant value in a linear function) parameters that are updated during the training phase of the model.  

The universal approximation theorem \citep{Hornik1989} states that \texttt{ANNs} with just a single layer can approximate any continuous function with a finite number of neurons. Here we train a simple Feed-Forward Neural Network (FFNN) \citep{Bebis1994} to map physical parameters to simulated disk-wind spectra ($y$), using both the \texttt{fast32} and \texttt{slow64} disk-winds. 

The inputs to the first layer are the parameters describing the AGN spectra $\mathbf{x}^0$ = 
$\{\Gamma,\,\mw,\,f_\nu,\,\lx,\,\mu , R_{\rm in} \}$ and the output of the final layer $\mathbf{x}^N$ are the predicted spectral values ($\hat{y}$). The trainable parameters ($\mathcal{NP}$) of the network are updated to optimise the loss function by comparing the predicted spectral values with the true spectral values. We explored the use of various loss functions and we found that the mean square error loss function,
\begin{equation}
    {\rm L2} = \sum_i (y_i - \hat{y}_i)^2,
\end{equation}
was most suited to this problem, as it is simple to compute and sensitive to outliers: an important characteristic to ensure absorption and emission lines are conserved. Furthermore, we experimented with the use of various activation functions\footnote{The activation function is a mathematical function that is added to an \ann in order to ensure non-linearity. In this way the \ann can learn the complex patterns of the training data `fed' into it.}: linear, tanh, exponential linear unit (ELU), and sigmoid \citep[see e.g.][]{Nwankpa2018}. Additionally, we tested the activation functions outlined in \cite{Alsing2020}, which was developed specifically to reproduce well both to smooth and sharp features -- again, an important feature for spectra. Nevertheless, we found that these activation functions underperformed compared to the rectified linear unit (ReLU) activation function on our data set,
\begin{equation}
    a(x) = \max(0, x). 
\end{equation}
This activation ensures that the outputs are positive, which is a key requirement for spectra. Under this same constraint, it is not possible to fit spectra in log units, where values can be $\le$\,0. In the case of log spectra, the network would have to be redesigned with some other activation function such as tanh, and/or a linear final activation function. Our emulator network consists only of fully connected layers, the best of which used 3 dense layers, each with 1000 neurons (\autoref{fig:Architecture}). In \ann a dense (or hidden) layer is located between inputs and outputs of the algorithm and performs non-linear transformations (i.e., fitting complex data) of the inputs and directs them into the outputs. They are referred to as dense (or hidden) because the `true' values of their neurons are unknown. In total, this results in $\mathcal{NP}=2,009,000$ trainable parameters (see Appendix\,\ref{sec:Network parameters} for the derivation).

The network was trained over 1500 epochs, where each epoch  comprises the entire data cycle, however for improved efficiency (i.e., not to feed the data at the same time), the parameters of the network are updated in batches. For the training, we use an incremental batch size updating from 1 to 100, to 1000 at each 500 epoch interval. The batch size, is the size of the training data sub-sample that is used to optimise the weights at a time. Larger batch sizes require more memory to load that can result in slower training, but smaller batch batch sizes give more stochastic loss which can will also take longer for the network to reach global minima. We use an increasing batch size that is equivalent to decreasing the learning rate \citep{Smith17}. The learning rate is another hyper-parameter that determines the size of the changes made to weights at each step. This aids the network in reaching the minimum loss, because as you approach the minima you need to make smaller changes to the weights or you will overshoot. Similarly, slowly increasing the batch size provides more confidence in the direction of descent to the minima as opposed to the stochastic descent provided by a small batch size. Additionally, early stopping \citep{Yao2007} is implemented to prevent over fitting. This ends the training process once the model is no longer improving. We use the adaptive optimiser Adam \citep{Kingma2014} to update the weights with learning rate of 10$^{-3}$. 

\begin{figure*}
    \centering
    \includegraphics[width=\linewidth]{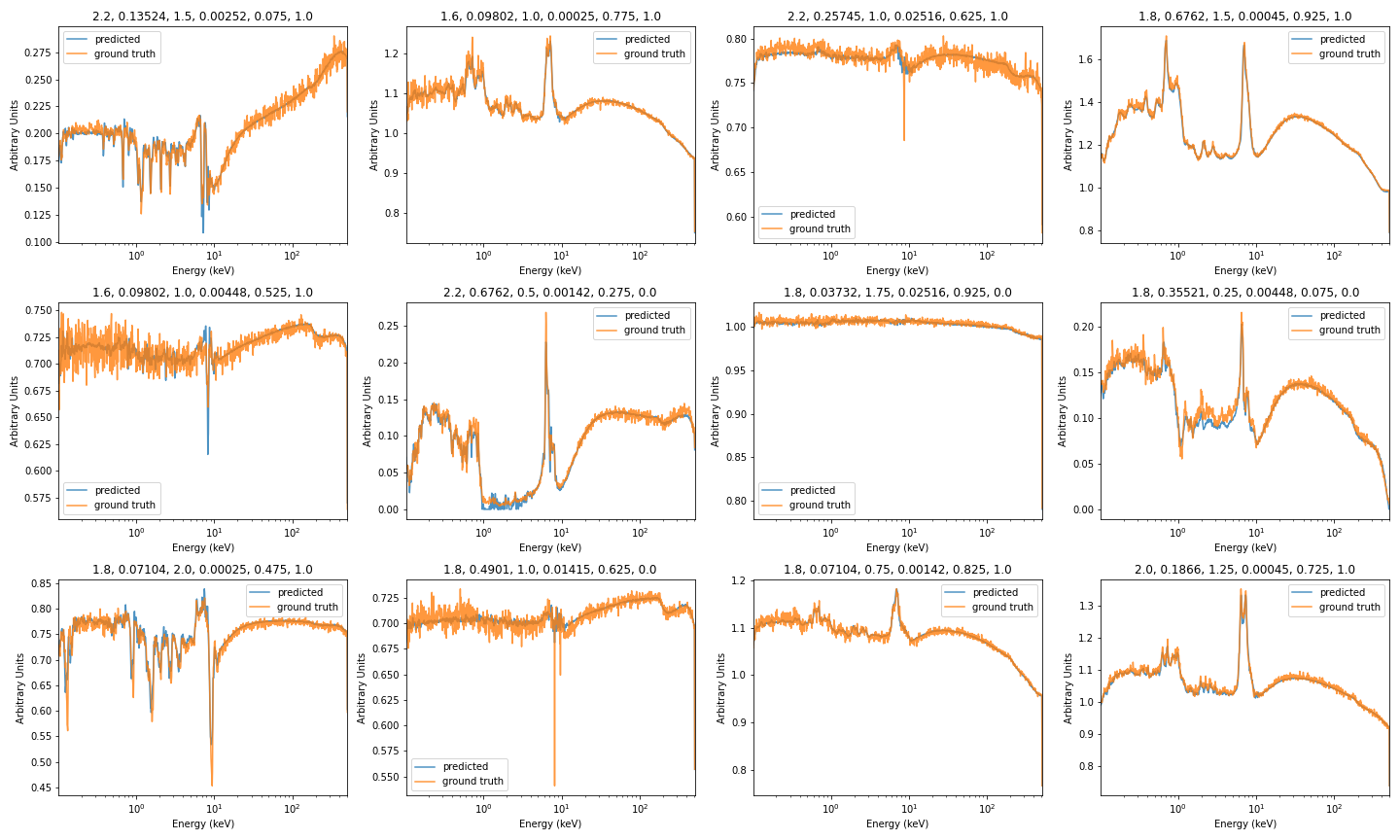}
    \caption{Randomly chosen examples of $12$ ground truth spectra (orange) out of the $17,280$ available from the test data ($10\%$ of total) with the corresponding emulated predicted spectra (blue). Note the \ann had not seen the spectra from the test set ($10\%$ of the total) during training. The input parameters corresponding to \{$\Gamma$, $\mw$, $\fv$, $\lx$, $\mu$, $\rmin=64\rg\,(0.0)\rm~or~32\rg\,(1.0)$\} are listed above each plot.}
    \label{fig:test_spectra}
\end{figure*}

In addition to training data that are used to optimize the network, additional data are required to validate the model, to ensure that it will generalise to new data. These data are seen during the training of the network to determine when to stop training. The performance of the trained network was then evaluated on additional test data that are \textit{not seen} during the training of the network. In total we have $172,800$ \mcrt synthetic spectra available for the \ann and we choose a train-validation-test split of $0.8$--$0.1$--$0.1$. This equates to $138,240$ spectra for training, $17,280$ for validation and $17,280$ for testing. The training set was checked to ensure a good representation of all parameters was included. 
The final L1 (absolute error),
\begin{equation}
{\rm L1} = \sum_{i} |y_i - \hat{y}_i|,
\end{equation}
and L2 (mean square error) loss on the validation data was 0.0071 and 0.0002, respectively. The L1 and L2 statistics for the test data set are 0.0071 and 0.0001, respectively. \autoref{fig:test_spectra} shows some examples of the spectra predicted by the emulator from the test data set.

\subsection{Mitigating interpolation issues with emulation}
\label{sub:Mitigating interpolation issues with emulation}

\begin{figure}
    \centering
    \includegraphics[width=8cm]{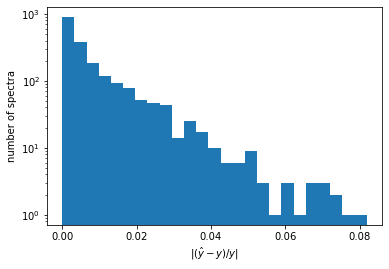}
    \caption{Histogram of absolute fractional error of the emulated spectra on the 2000 additional simulations.}
    \label{fig:accuracies}
\end{figure}

\begin{figure}
    \centering
    \includegraphics[width=8cm]{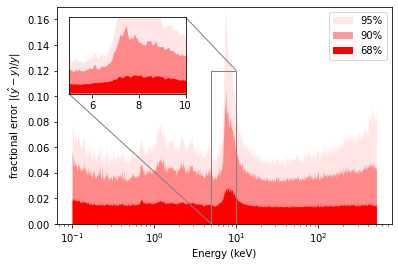}
    \caption{Fractional error on flux at different energies for the 2000 additional test simulations from $0.1$--$511\kev$. We show 68$\%$, 90$\%$, and 95$\%$ of the sample. }
    \label{fig:frac_err_den}
\end{figure}

Until now, the data used to train and test the network are \mcrt simulations from 2 grids of parameters. But we need to know if the emulator is capable of reproducing parameter values between all the grid points. To do this, the trained network is further tested against $2000$ new \mcrt simulations, where $100$ of each of the parameters are drawn from uniform distributions: $\Gamma \sim \mathcal{U}(1.6,2.4)$, $\mw \sim \mathcal{U}(0.0196, 0.6762)$, $f_\nu \sim \mathcal{U}(0.25,2)$, $\lx \sim \mathcal{U}(2.52\times10^{-4}, 2.52\times10^{-2})$, and the launch radius from a binomial distribution $R_{\rm in} \sim \mathcal{B}(1,0.5)$, corresponding to the \texttt{fast32} and \texttt{slow64} winds. For each spectrum, we have corresponding $\mu$ values of 0.025 to 0.975 in steps of 0.05. \autoref{fig:accuracies} shows the fractional offset of the predicted from the ground truth spectra,
\begin{equation}
    {\rm fractional\, error} = \left|\frac{\hat{y}-y}{y}\right|.
\end{equation}
The fractional error is in most cases smaller than the noise on the simulated spectra, and we find no bias with respect to any particular parameter. Typical errors are of per-cent level across the entire energy range (\autoref{fig:frac_err_den}), although a non-negligible error is seen in the $7$--$8\kev$ band, corresponding to \fexxvxxvi transitions in both emission and absorption. 

\begin{figure}
    \centering
    \includegraphics[width=\linewidth]{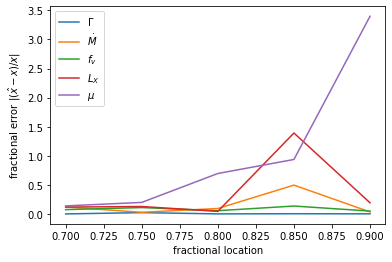}
    \caption{The fractional error on individual parameters fit using the \mcrt tables based on \textbf{5 different} spectra generated from \xrade with increasing fractional error at the $8\kev$ band \textbf{(x-axis)}.}
    \label{fig:ind_error_8keV}
\end{figure}

To investigate the influence of the fractional error in the 7-8 keV band on the parameters (\autoref{fig:frac_err_den}), we take take the fractional error on the flux values at from the test simulations at 8 keV. We order the error values and take the parameter set corresponding to the 70$\%$, 75$\%$, 80$\%$, 85$\%$ and 90$\%$ error value as shown in \autoref{fig:ind_error_8keV}. From these 5 parameters sets we emulate spectra using \xrade and create CCD observations, by using the 
\xmm EPIC-pn response and background files corresponding to the \pds ObsCD observation in 2013 (see \autoref{sec:application on pds456}), between the $0.3$--$10\kev$ energy range (i.e., the \xmm band-pass), using \xspec. The observation is then fit using the \mcrt tables. The parameters are generally well recovered despite the differences between the \mcrt tables and \xrade. We find that the recovery of $\bm{\mu}$ is the only parameter that is affected by the uncertainty on the Iron K band pass. This parameter is the one that affects the shape of the spectral features the most (see Appendix\,\ref{subapp:Influence of geometry in disk-wind features}).

The fractional error seems to increase, not as severely, at energies corresponding to other wind features e.g., O\textsc{viii}, Ne\textsc{x}, and Si\textsc{xiv}. The network could benefit from training data with more spectral points in these regions, and ideally a more finely sampled grid of parameters (see \autoref{sec:conclusion and future work}). In particular, in the near future, we also aim to re-generate steps for the \dwf and \dws grids in linear (rather than log) space for the $\lx$ and $\mw$ parameters. This will likely increase the accuracy of mapping these parameters through the emulator and this could be especially important in training the emulator at the higher $\mw$ range, which is currently more sparsely sampled in logarithmic space. As a consequence, this may also reduce the fractional error seen over the iron\,K band-pass in \autoref{fig:frac_err_den}.

\begin{figure*}
  \includegraphics[width=\linewidth]{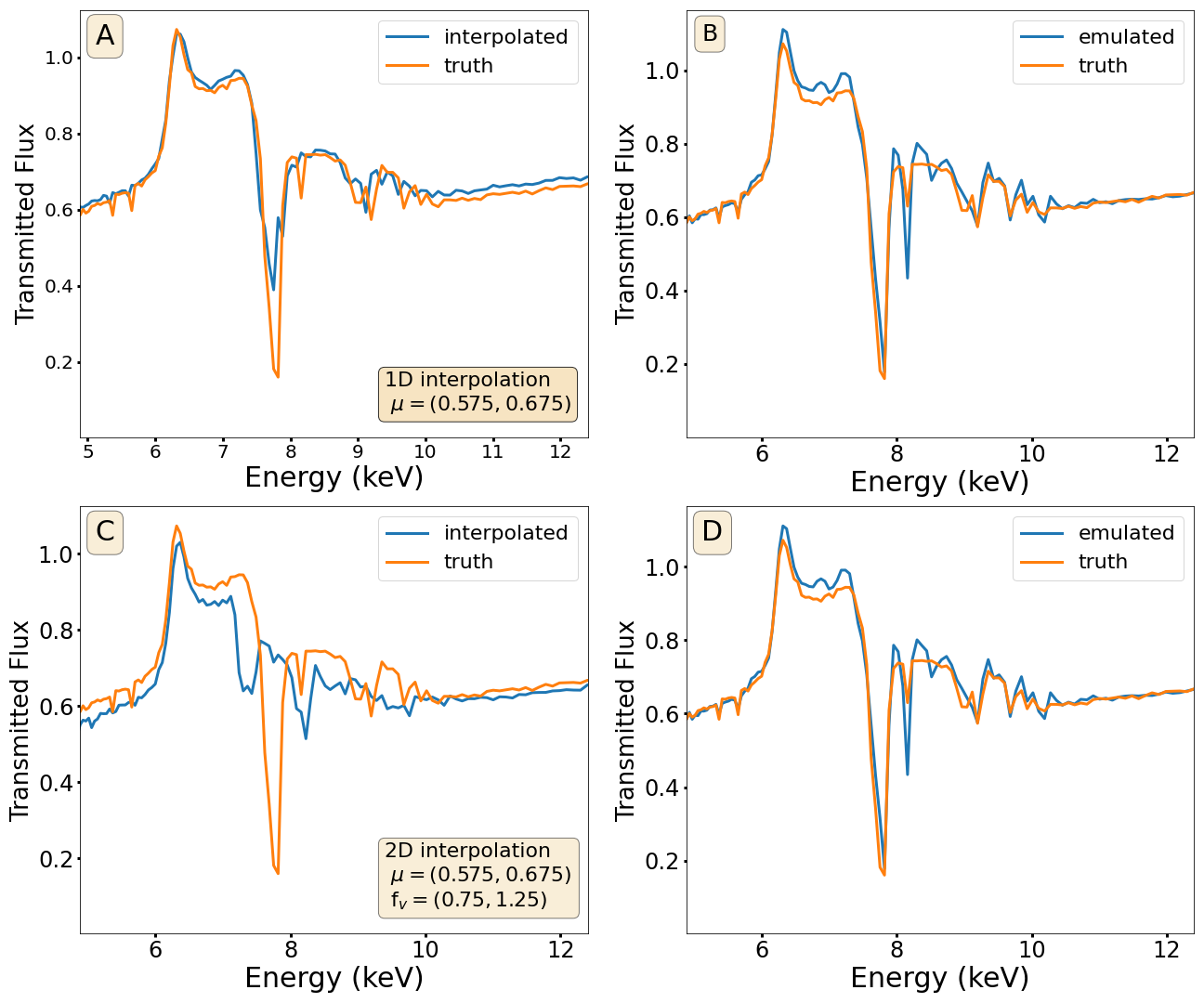}
  \caption{A comparison between interpolated spectra, normally produced in \xspec, and those predicted by our trained \ann. \textit{Top row:} in panel\,A we show the true spectrum of $\mathbf{x}_{\rm true} =[\Gamma=2.0,\mw=0.257,\fv=1.0,\lx=2.52\times10^{-4},\mu=0.625,\rmin=64\rg]$ (orange) and the spectral prediction using interpolation between $\mu=(0.675,0.575)$. Same in panel\,B, but the true spectrum is predicted by the emulator (blue). Panel\,C: same as panel\,A but with a 2D parameter interpolation involving $\mu$ and $\fv$ (see text box). The resulting interpolated spectrum is not able to accurately recover the amplitude and shape of the spectral lines. Panel\,D: the emulator is able to produce an accurate mapping of the ground truth spectrum.}
\label{fig:1d_2d_int_vs_emu}
\end{figure*}

This test demonstrates that not only are we able to use the emulator for parameter values within the training range, but also on parameter values that lie between points on the simulated grid. The trained emulator can predict spectra for a particular parameter set in $\sim$\,0.04 seconds in comparison to $\sim$2--3 hours when using the \mcrt pipeline, which allows us to emulate finer grids of models more efficiently. In this light, we test whether the predictions of our emulation process are able to reproduce a true (or ground) spectrum. We then compare them with spectra arising from standard interpolation between grid values, which is normally occurring in X-ray fitting packages such as \xspec. For this test, a true spectrum can be selected from any of the $17,280$ available test spectra. We chose two test cases in \autoref{fig:1d_2d_int_vs_emu}, one (upper panels) where there is one free wind parameter ($\mu$) and one where two parameters ($\mu$ and $\fv$) are varied (\autoref{fig:1d_2d_int_vs_emu}, lower).

For the 1D test we considered the case where the true spectrum has the following parameters: $\mathbf{x}_{\rm true} =[\Gamma=2.0,\mw=0.257,\fv=1.0,\lx=2.52\times10^{-4},\mu=0.625,\rmin=64\rg]$. Here the 1D (panel\,A, upper-left \autoref{fig:1d_2d_int_vs_emu}) interpolation (blue) between $\mu=0.575$ and $\mu=0.675$ (to order to reproduce a real value of $\mu=0.625$). Here the interpolation does a reasonable job in reproducing the true spectrum with the following conditions: $\mathbf{x}_{\rm true} =[\Gamma=2.0,\mw=0.257,\fv=1.0,\lx=2.52\times10^{-4},\mu=0.625,\rmin=64\rg]$ (orange), but underestimates the profile depth. The emulated true spectrum plotted in panel\,B (blue) is better at reproducing the depth of the absorption trough at $\sim8\kev$, but slightly worse at estimating the higher-order transition at $\sim8.3\kev$. Overall, both methods reasonably predict the true spectrum between $5$--$12\kev$.

In examples C and D we compare a 2D interpolation (i.e., two parameters of interest) between $\mu=(0.575,0.675)$ and $\fv=(0.75,1.25)$ to respectively reproduce $\mu=0.625$ and $\fv=1.0$. In this scenario it is much harder for the interpolated spectrum to reproduce $\mathbf{x}_{\rm true}$ as these parameters produce both a shift in energy and depth simultaneously. Clearly the interpolated spectrum fails to reproduce the true spectrum. On the other hand the emulated spectrum is a closer match to $\mathbf{x}_{\rm true}$ (panel\,D). It is worth noting that any interpolation issues in $\mw$ nor $\lx$ are not as dramatic, given that they are mainly affecting the profile depth and do not tend to produce a shift in energy between the points.



\begin{figure}
  \includegraphics[width=\linewidth]{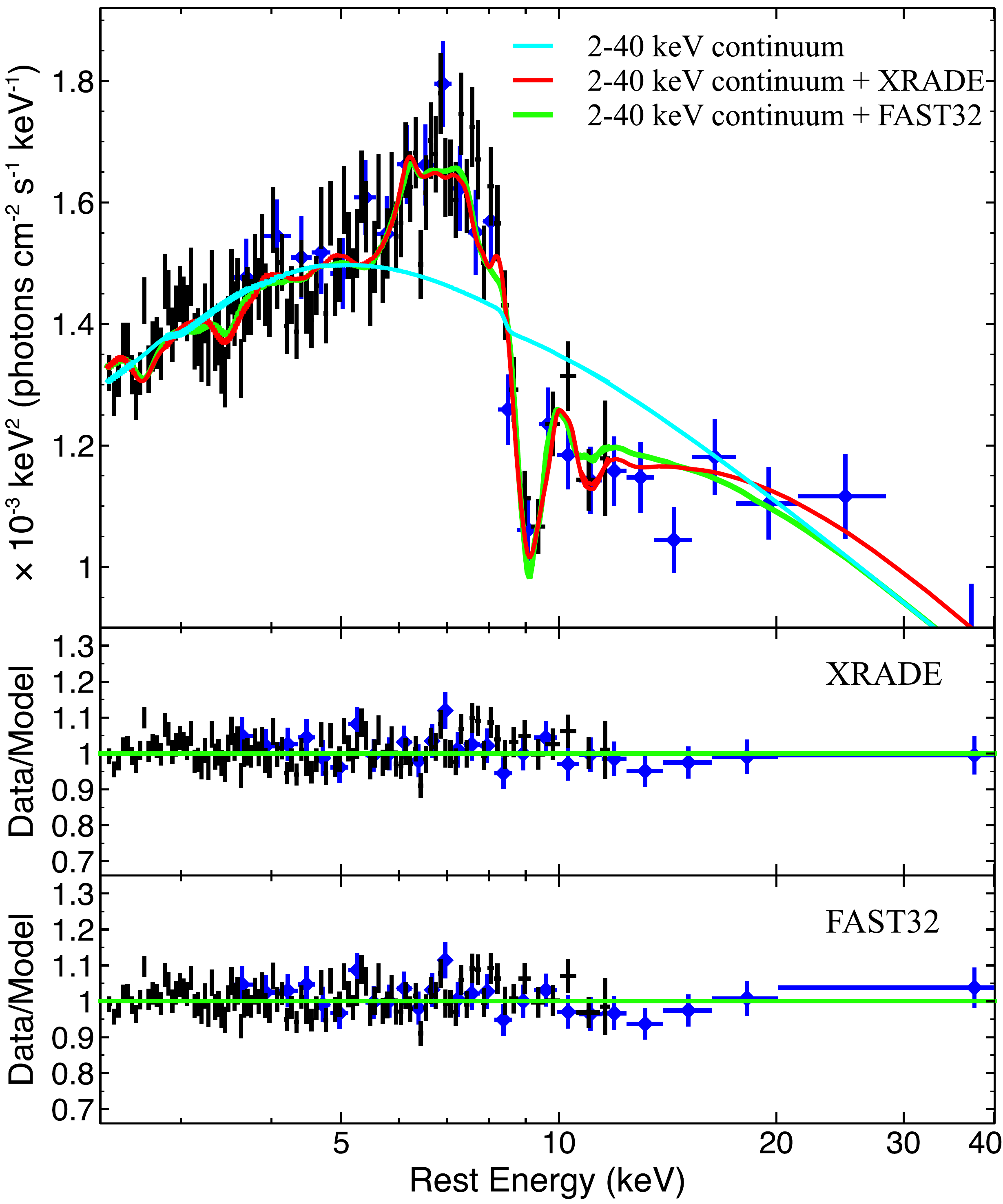}
  \caption{Top: Unfolded \xmm (black) and \nustar (blue) spectra (ObsCD) of \pds (against a $\Gamma=2$ power-law) between $2$--$40\kev$ with the continuum only model (cyan), \xrade (red) and \dwf (green) superimposed. Bottom: The corresponding data/model ratio plots. Both models do an excellent job in fitting the P-Cygni feature, of which they self-consistently fit the broad emission (scattered/reflected component) and absorption (direct component).}
	\label{fig:pcygni_emu}
\end{figure}

\section{Observational data: Fitting the powerful disk-wind in \pds with \textit{XRADE} and \textit{FAST32}}
\label{sec:application on pds456}

The generated \xrade spectra are tabulated into \textsc{fits} files and can be used as multiplicative grids within \xspec. In this section we want to compare the overall performance and reliability consistency check of our \mcrt and \xrade tables with real CCD data. we also want to check and compare them in the the ability predicting values between the grid points. 

As a test case we consider the `prototypical' (and most studied) disk-wind hosted in the luminous quasar \pds. A large monitoring campaign, covering 6 months, was carried out between 2013 and 2014 and consisted of five joint \xmmnu observations (ObsA--ObsE) of $\sim100\ks$ each. During these observations, a prominent and persistent P-Cygni profile was revealed (N15). Such a feature is characterized by the combination of a broad emission and absorption profile, where the former is produced by scattered photons off the wind averaged from all angles and the latter from transmitted photons through the material. ObsC and ObsD were separated by only $\sim$\,3 days, so their spectra were virtually identical. As per N15, we subsequently combined them into a single ObsCD observation resulting into a total net exposure time of 195 ks, showing a P-Cygni feature of unprecedented quality. The \xmmnu data considered here are the EPIC-pn \citep{Struder01} and FPMA+FPMB \citep{Harrison13}, respectively and they are reduced following the procedure presented in N15.

From what was discussed in \autoref{sub:Launch Radius}, the initial setting of $\rmin$ has a direct impact on the range of outflow velocities that can be measured (see \autoref{eq:vinf_range_limits}). In this paper we chose \dwf for our comparison with \xrade. Note that \dwf, was initially generated based on the range of velocities observed in \pds \citep[][e.g., $\vout/c=0.25-0.35$]{Matzeu17b,Reeves18PDS} since its first detection with \xmm in 2001 \citep{Reeves03}.  On the other hand, by following the same prescription in \autoref{sub:Launch Radius}, \dws was successfully applied in modelling the powerful disk-wind observed in the Seyfert\,2 galaxy \mcg \citep{Braito22}.

In \autoref{fig:pcygni_emu}, we show the unfolded \xmmnu spectra  of \pds (ObsCD) between $2$--$40\kev$ against a simple $\Gamma=2$ power-law. Once the continuum (cyan) is accounted for, there are strong residuals in the \fe region that correspond to the P-Cygni feature. From a visual inspection the centroid energies are located at $E_{\rm rest,em}\sim7\kev$ and $E_{\rm rest,abs}\sim9\kev$ for the emission and absorption component, respectively. 
The model in \xspec is expressed as:  
 
\begin{equation}
\texttt{Tbabs} \times \texttt{pcfabs} \times (\texttt{powerlaw} \times \texttt{highecut}) \times \xrade~({\rm or}~\dwf),
\label{eq:best-fit model}
\end{equation}

\noindent where \texttt{Tbabs} is the Galactic absorption of $\nhgal=2.9\times10^{21}\cmsq$ \citep{Reeves21}. To model the soft X-ray spectral curvature we adopt a layer of neutral partial covering (\texttt{pcfabs} in \xspec) with $\nh=7.9_{-2.5}^{+2.1}\times10^{22}\cmsq$, and covering fraction of $\mathcal{C}_{\rm frac}=0.37_{-0.02}^{+0.05}$. A high-energy rollover (\texttt{highecut}) fixed at $E_{\rm cut}=100\kev$ was also adopted and a cross-normalization factor between the \xmm and \nustar detectors was measure at $\mathcal{C}_{\rm cal}=1.10\pm0.02$. For this test, we generated a customized \xrade grid with the values tabulated in \autoref{tab:xrade_values}. 

\begin{table}
  \caption{Customized \xrade model values and ranges used for \pds ObsCD.}
\centering
  \begin{tabular}{l|c c c}
\hline

Parameter        &value range                        &$\Delta$ value   &Steps                   \\

\hline

$\Gamma$         &$1.6$--$2.4$                       &$0.1$            &$9$                    \\

$\mw$            &$0.05$--$0.65$                     &$0.05$           &$13$                     \\

$\fv$            &$0.25$--$2.0$                      &$0.097$             &$19$\\

$\lx$            & (0.05--1.5)\,$\times$10$^{-2}$                      &9.7$\times$10$^{-4}$             &$16$                            \\

$\mu$ &$0.2$--$0.9$                      &$0.05$             &$15$                             \\

$\rmin$        &$32\rg$                              &--                 &--                      \\

\hline

\multicolumn{4}{c}{Number of emulated spectra: $533,520$}\\ 

\hline

  \end{tabular}
\label{tab:xrade_values}
\end{table}

Fitting the P-Cygni profile with \xrade yielded a mass outflow rate of $\mw=0.318_{-0.046}^{+0.014}$, i.e. about $30\%$ of $\medd$. In \pds, with $\mbh\sim10^{9}\Msun$ and $\ledd\sim1.3\times10^{47}\ergs$, then $\medd\sim40\,\Msun\,{\rm yr}^{-1}$ for $\eta=0.06$, $\mout\sim10\,\Msun\,\rm yr^{-1}$. The X-ray ionizing luminosity is $\lx=0.272_{-0.061}^{+0.090}\times10^{-2}$ or $\sim0.3\%$ of $\ledd$ , i.e., $L_{2-10\kev}\sim4.0\times10^{44}\ergs$. By comparison, the directly observed intrinsic $2$--$10$ keV luminosity is of the order of $\sim5\times10^{44}\ergs$ and hence consistent with the \xrade predicted value, see \autoref{tab:pdsCD_xrade_values}. A \los orientation angle of $\theta\sim50^{\circ}$ (i.e., $\mu=0.63_{-0.02}^{+0.01}$) with respect to the polar axis is required, suggesting that the sight-line fully intercepts the innermost and fastest wind streamline, hence explaining the prominence (and high degree of blueshift) of the P-Cygni feature. The terminal velocity parameter was measured at $\fv=1.33_{-0.04}^{+0.03}$ and, as \xrade was generated by assuming a launch radius of $\rmin=32\rg$, this translates into a terminal wind velocity of $\vinf =-0.33_{-0.01}^{+0.01}c$. 

Note that the input photon index of the \xrade model is tied to the \texttt{powerlaw} continuum at $\Gamma=2.19_{-0.02}^{+0.05}$. The addition of \xrade resulted in a large improvement on the fit statistics by $\dchidof=-231.4/4\,(>99.99\%)$, for an overall best-fit $\chidof=655.3/682$. We subsequently replaced \xrade with our \mcrt generated \dwf in \autoref{eq:best-fit model}. We find that both fits are excellent and almost identical with $\chidof=659.7/682$ (see \autoref{fig:pcygni_emu}\,(bottom right). During the fitting procedure in \xspec, the `delta' value parameter has been set to be 0.001 (via the \textsc{xset} command) so that a like-for-like comparison could have been achieved between \xrade and \dwf. Moreover, the same best-fit values were returned when restoring the original fixed delta values of the model (i.e., via the command \textsc{xset} delta 0.0)

The values are largely consistent with \xrade, as shown in \autoref{tab:pdsCD_xrade_values}. This initial consistency test demonstrates that both physical models provide an excellent fit to the P-Cygni like profile in \pds and that \xrade is able to reproduce the results obtained by the \mcrt grid. Note that errors measured in both grids are indeed similar due to CCD spectral resolution of the data which illustrates that, at the resolution of the data, \xspec interpolation upon the \mcrt table models achieves an equally adequate parameterization of the data as per the emulated \xrade tables. However the limitations of the former and over-reliance of interpolation is more likely to have a significant impact for calorimeter resolution spectra, which we further discuss below.

\begin{figure}
\centering
\includegraphics[width=\linewidth]{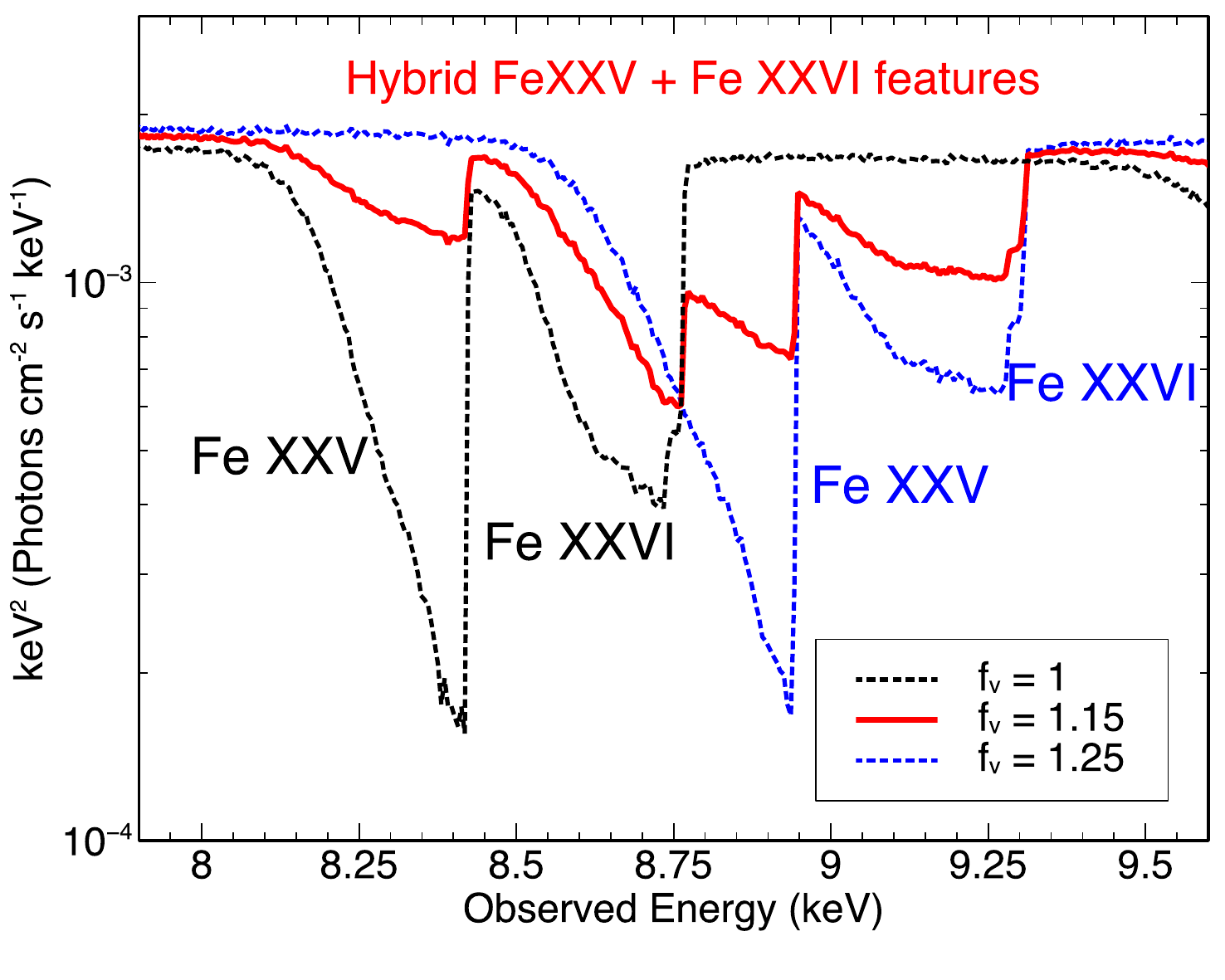}
\caption{Three simulated micro-calorimeter resolution (i.e., $2\ev$ at 6.4\,keV) spectra, using an high resolution \dwf grid, corresponding to $\fv=1$\,(black) and $\fv=1.25$\,(blue). The remaining disk-wind parameters are fixed to $\Gamma=2$, $\mw=0.3$, $\lx=0.5$ and $\mu=0.625$. The interpolated spectrum between $\fv=1$ and $\fv=1.25$, corresponding to $\fv=1.15$ is shown in red. Note the asymmetry of the profiles visible at high resolution is a direct result of the acceleration of the gas along the streamline.}
\label{fig:dw_hires_interpolation}
\end{figure}

In \autoref{fig:dw_hires_interpolation} we show three simulated \athenaifu resolution (i.e., $2\ev$ at 6.4\,keV) spectra using the high-resolution disk-wind grid (\texttt{f32hires}; see \citealt{Parker22DWDEGENERACY} for details). \texttt{f32hires} was a \mcrt generated table to match the micro-calorimeter resolution data of \xrismres and \athenaifu with a total of 10,000 energy bins (i.e., with an energy resolution of $\Delta E=1.8\ev$) between 0.1--20\,keV. Due to its high CPU cost, \texttt{f32hires} is in a preliminary stage and is limited to 2400 grid points, however it will be expanded in the near future.    

Here we keep all the parameters fixed (see caption) whilst the changing the velocity factor parameter to $\fv=1$\,(black), $\fv=1.25$\,(blue) and the interpolated value of $\fv=1.15$\,(red) between the former two grid points. As expected, both the highly ionized (i.e., \fexxvabs and \fexxviabs) absorption features are prominent in both $\fv=1$ and $\fv=1.25$ spectrum, although more blueshifted in the latter. The intermediate (interpolated) point seems to generated a spectrum that is characterized by some hybrid set of absorption feature caused by interpolation. The intermediate (interpolated) spectrum at $\fv=1.15$ is characterized by a hybrid set of absorption features caused by interpolation in energy space, between the $\fv=1$ and $\fv=1.25$ grid points. In fact such an issue is already striking, unlike in the CCD resolution framework (see \autoref{fig:1d_2d_int_vs_emu}), in the simplest 1D interpolation discussed in \autoref{sub:Mitigating interpolation issues with emulation}. A more detailed set of experiments will be performed and reported on a following companion paper.

At this stage, the key contrast between these two tables is the vast difference in the CPU time required to generate these grids. In fact, to produce the $86,400$ synthetic spectra in \dwf required an overall CPU time of $\sim$\,4 months on $600$ cores at 50\,Gb RAM (per core), against an impressive time-scale of $\sim$\,4 seconds for generating $533,520$ emulated spectra for the \xrade table. Note that our emulator has the flexibility to generate parameter ranges with unprecedented resolutions within minutes.

\begin{table}
  \caption{\xrade and \dwf model results for \pds ObsCD. The uncertainties are calculated at a $90\%$ confidence level which correspond to a $\dchidof=2.71/1$. The power-law normalization is in unit of $\rm photons\,keV^{-1} cm^{-2}\,s^{-1}$ at 1\,keV }
\centering
  \begin{tabular}{l | c c}
\hline

Parameter          &\xrade                             &\dwf                \\

\hline

$\Gamma$           &$2.19_{-0.05}^{+0.02}$             &$2.21_{-0.04}^{+0.05}$ \\

$\mw$              &$0.318_{-0.046}^{+0.014}$          &$0.243_{-0.131}^{+0.056}$ \\

$\fv\,\left( \vinf \right)$              &$1.33_{-0.04}^{+0.03}\,\left(-0.33_{-0.01}^{+0.01}c\right)$             &$1.31_{-0.04}^{+0.08}\,\left(-0.33_{-0.01}^{+0.02}c\right)$  \\

$\lx$              &$0.273_{-0.061}^{+0.090}\,\times10^{-2}$  &$0.208_{-0.130}^{+0.058}\,\times10^{-2}$ \\

$\mu=\cos\theta$   &$0.629_{-0.017}^{+0.010}$                 &$0.652_{-0.022}^{+0.008}$  \\

\hline

$\lognh$                    &$22.8_{-0.4}^{+0.4}$             &$22.9_{-0.4}^{+0.4}$ \\

$\mathcal{C}_{\rm frac}$   &$0.39_{-0.04}^{+0.04}$            &$0.37_{-0.02}^{+0.03}$  \\

$\rm norm/10^{-3}$         &$3.2_{-0.3}^{+0.05}$              &$2.7_{-0.2}^{+0.3}$ \\

$C_{\rm cal}$              &$1.10_{-0.02}^{+0.02}$            &$1.10_{-0.02}^{+0.02}$ \\

\hline

 $\dchis$ & $655.3/682$ & $659.7/682$\\

\hline

  \end{tabular}
	\label{tab:pdsCD_xrade_values}
\end{table}

\subsection{Global parameter exploration}
\label{subsec:bxa}
We sought to test the emulated parameter space created with \xrade via global parameter exploration. We use the same \textit{XMM-Newton}/EPIC-pn and \textit{NuSTAR} \pds datasets as described in \autoref{sec:application on pds456} and an identical model setup. For the purposes of comparing the different parameter spaces, we use the \dwf and \xrade table model as in \autoref{eq:best-fit model} (see \autoref{tab:xrade_values}). We employ the Bayesian X-ray Analysis (\textsc{bxa} v2.10; \citealt{Buchner14}) software platform which connects the nested sampling algorithm \texttt{MultiNest} \citep{Feroz09} with the Xspec fitting environment. In brief, nested sampling (see \citealt{Buchner21NS} for a recent review) stores a set of parameter vectors drawn from the prior distribution. The lowest Likelihood parameter vector is iteratively replaced with a new one of higher Likelihood, until some termination condition is met. In this way, the algorithm scans the global prior-defined parameter space and is thus a useful tool for visually exploring and comparing the multi-dimensional parameter spaces associated with \dwf and \xrade.

We assign uniform priors to all parameters apart from the partial covering absorber column density and intrinsic power-law normalisation which were assigned log-uniform priors, and the multiplicative cross-calibration constant which was assigned a custom log-Gaussian prior with mean zero (i.e. a linear cross-calibration of unity) and 0.1 standard deviation. This choice of prior is useful for the cross-calibration to avoid negative values, whilst also peaking close to unity (e.g., \citealt{Madsen17}). The same 10 free parameters were used in both models.

\begin{figure*}
  \includegraphics[scale=0.11]{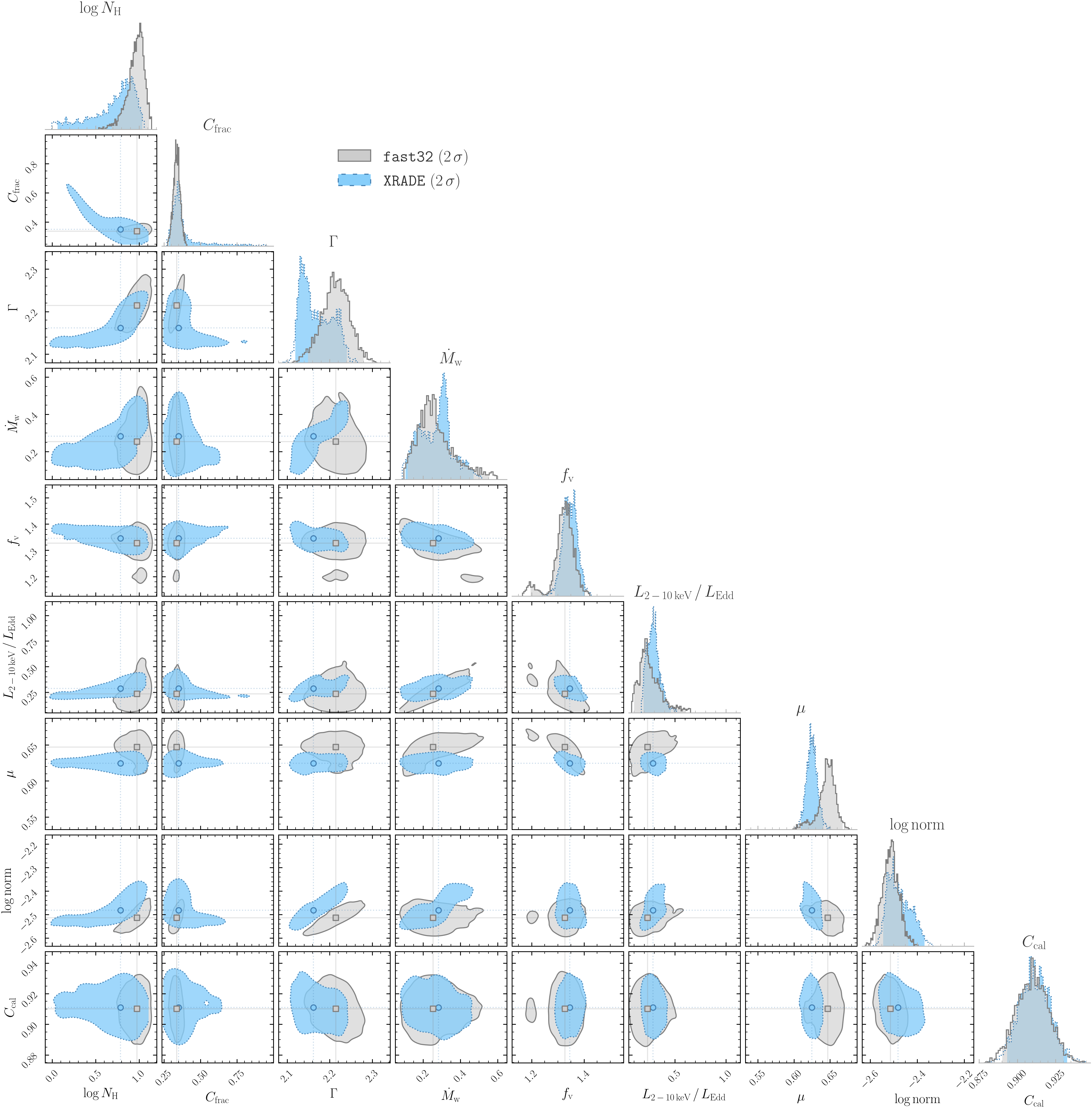}
  \caption{Corner plot showing the results obtained from fitting the \textit{XMM-Newton} and \textit{NuSTAR} spectra with \dwf (grey) and \xrade (blue) using \bxa. Shaded regions show the 2\,$\sigma$ confidence level.}
	\label{fig:bxa_pds456_corner}
\end{figure*}

The result of the fits are shown in \autoref{fig:bxa_pds456_corner} with grey and blue contours for \dwf and emulated \xrade tables, respectively. Shaded regions represent the $2\sigma$ level, though note that the percentage of points encompassed by the 2D contours is not the same as in the 1D histograms\footnote{See \href{https://corner.readthedocs.io/en/latest/pages/sigmas.html}{https://corner.readthedocs.io/en/latest/pages/sigmas.html}}. In general, the parameter space attained with \xrade appears to match the \dwf parameter space well with good agreement within 2$\sigma$. The majority of individual posterior shapes also show good agreement, indicating that the emulation process is able to  reliably map different regions of parameter space to spectral space.

There are some parameters that have different posterior shapes, e.g., $\mu$. Disagreements between posterior shapes could indicate that particular regions of the emulated spectral/parameter space require more training data as input. Alternatively, even though both models were fit with \xspec, the emulated parameter grid of \xrade was finer than \dwf, hence with the corresponding interpolation between adjacent grid points performed over smaller parameter steps with \xspec. We note that the higher-resolution \xrade table does not necessarily mean that the confidence intervals should be smaller, since the aim of the emulator is to reproduce the multi-dimensional parameter space associated with the original \dwf model as accurately as possible. The ultimate limitation to the confidence intervals is thus the data quality, since the emulated \xrade model was trained on \dwf originally.

If the input training data was sufficient for the \ann to learn the complex mapping process involved, posterior differences could hint to alternative parameter estimation with emulation vs. interpolation. However, since $\mw$, $\fv$ and $\mu$ have a very strong (and/or non-linear) relation to the observed spectral shape of the model, such parameters are most likely to suffer from interpolation issues, suggesting such parameters may require finer parameter resolution training grids in particular. Nonetheless, testing future emulated \xrade tables on real data with \bxa may be an efficient method to iteratively explore and check the emulated parameter space in detail.

\begin{figure*}
  \includegraphics[scale=0.7]{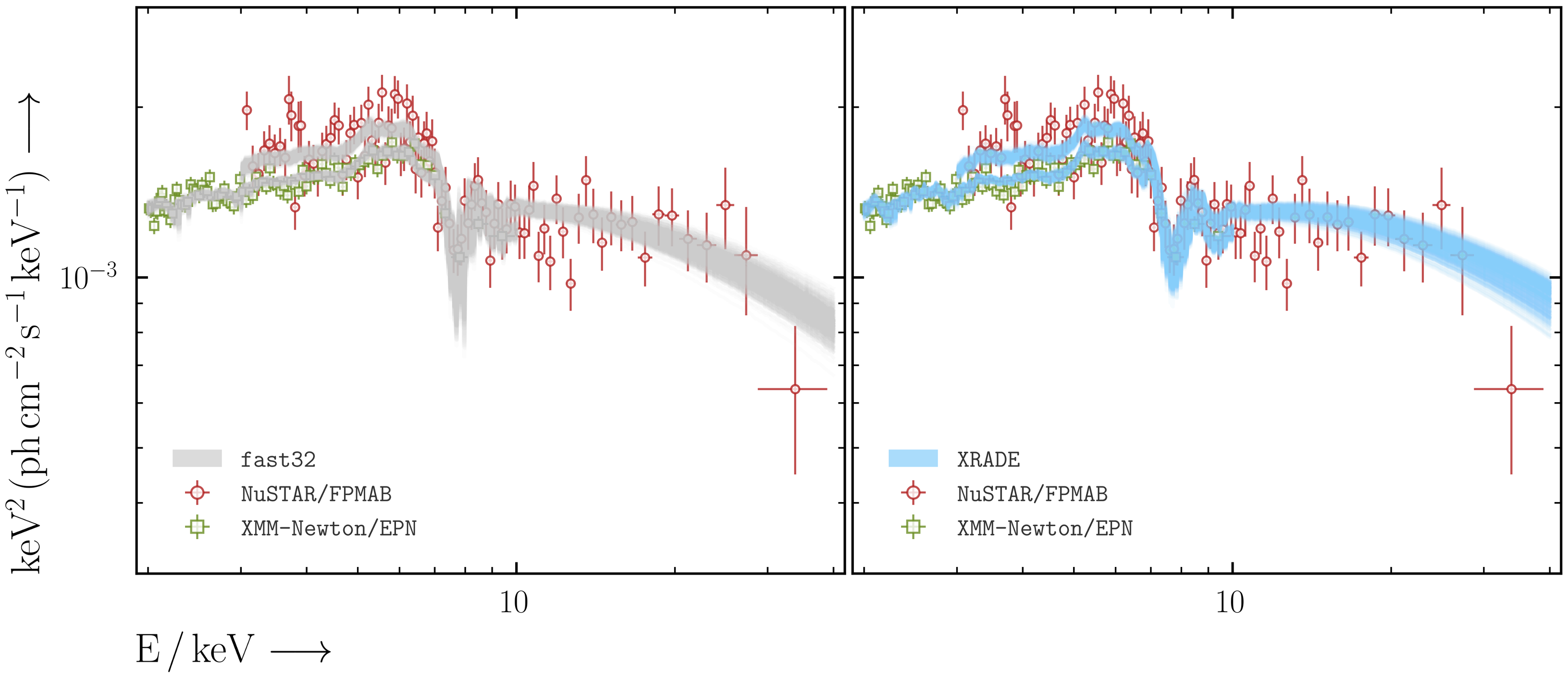}
  \caption{Unfolded model realisations from the \bxa fits with the original \dwf (grey) and \xrade (blue) multiplicative table models. The data adopted here are the same as per \autoref{fig:pcygni_emu}.}
	\label{fig:bxa_pds456_realisations}
\end{figure*}

\autoref{fig:bxa_pds456_realisations} presents an alternative comparison between the spectral fits performed with BXA in Section~\ref{subsec:bxa}. A total of 500 posterior parameter vectors from the \texttt{fast32} (left) and \texttt{XRADE} (right) model fits were loaded and over plotted with the unfolded spectral data. The models found for each dataset (distinguished by the cross-calibration) are plotted with the same colour in each panel and shaded regions represent the overall 500 realizations. Clearly the spectral shapes are very similar apart from a small difference at $\sim$8\,keV, in agreement with Figure~\ref{fig:frac_err_den}.

\subsection{Other models}
\label{sub:other_modles}
A model similar to \xrade is defined and used in \citet{Hagino15} (\textit{MONACO}: MONte Carlo simulation for Astrophysics and COsmology), which is then subsequently applied in \citet{Hagino16b,Hagino16a} to fit the disk winds in \pds, 1H0707--495 and APM\,08279$+$5255. Here, the same bi-conical structure is used  (see fig. 3 in \citealt{Hagino15}). \textit{MONACO} separates the wind structure into shells and then performs a series of \xstar runs to ascertain the ionization balance and the luminosity leaving and entering each layer. The radiative transfer is then performed using the He-like and H-like iron and nickel transitions along with Compton scattering. This has the benefit of being less computationally expensive than our disk-wind code, as the higher the number of lines which are tracked, the more computationally intensive the simulation. Therefore, the limited number of transitions allows a quicker exploration of the parameter space. The argument for only tracking the highly ionized species is that high-velocity winds are typically highly ionized. 

However, lower ionization species can survive in thicker winds and should be considered in a more general case. These lower ionization species may be observed at lower energies, such as the lower ionization lines observed in the \xmmrgs data of many AGNs. In PG\,1211$+$143 \citep{Pounds16,Reeves18PG1211} and \pds \citep{Reeves16,Reeves20rgs} these soft features appear to be physically associated with the highly ionized outflow. These features may be studied in more detail in the future by lowering the ionization in runs. This can be done by either lowering the source luminosity or increasing density through clumps within the streamlines. 

It is thus important to stress that the faster winds will not just produce more highly blue-shifted lines, but also produce intrinsically broader line profiles, both in emission and absorption. While in principle such profiles may be accounted for in other non-wind scenarios (e.g. by absorption through a co-rotating disk atmosphere, \citealt{Gallo11,Gallo13,Fabian20}), in \autoref{sec:application on pds456} we demonstrate that the broad P-Cygni like profile in \pds can be self consistently modelled by our Solar abundance, \dwf and \xrade table of models.

\section{Conclusions and Future work}
\label{sec:conclusion and future work}

In this paper we presented an improved version of the state-of-the-art disk-wind model obtained from a Monte Carlo multi-dimensional (2.5D) radiative transfer code initially developed by \citet{Sim08,Sim10}. For this purpose, we generated two large \mcrt tables, \dws and \dwf, of $172,800$ synthetic spectra, covering a much wider parameter space (see \autoref{tab:in_parameters}) than previously presented (S08; S10, \citealt{Reeves14,ReevesBraito19}). These will allow us to explore the physical conditions that characterize the accretion disk-winds across a wide range of sources, as our measurements are black hole mass invariant. As mentioned above, \dws has been already applied to \mcg \citep{Braito22}, and the \dwf will be applied to all the \pds data from 2001-2019 (Reeves et al. in prep).

We also presented the development and implementation of a novel emulator based on a purposely built \ann: X-Ray Accretion Disk-wind Emulator (\xrade). The method developed here works as follows. From the available \mcrt generated spectra, we fed $80\%$ (or $138,240$ spectra) into the \ann. A further $10\%\,(17,280)$ are used for validation and the remaining $10\%$ are exclusively kept for testing the emulated spectra. Our emulator is not only able to reproduce the \dws and \dwf synthetic spectra, which required a total of $\sim$\,8 months ($600$ cores) to be generated, but also to emulate $533,520$ spectra (see \autoref{tab:xrade_values}) within a $4$-minute timescale, i.e. $\sim$\,5 orders of magnitude faster, with an average mean square error of just $1.4\%$. 

After the training and validation process, our built \ann can emulate synthetic \mcrt spectra well within $10\%$ accuracy. As far as using \xrade in \xspec, we are able to successfully produce finer tables than \dws and \dwf as long as they are within the parameter boundaries set in the \mcrt tables. Any user can easily build a fully customized \xrade multiplicative table that will be suitable for spectral analysis in \xspec. A future test is, however, to explore whether a coarser and wider parameter grid can be used in order to localize regions of the parameter space to an acceptable level of precision, via e.g. the \bxa process and error searches. Once the parameter space is mapped, then finer grids can be adopted. 

We note that a finer \xrade table would still be susceptible to \xspec interpolation issues. Our foreseeable goal is to exploit the \ann impressive emulation rate to be \textit{directly} implemented in the fitting procedure. We aim at eventually bypassing interpolation based fitting programs such as \xspec, as well as grid development, and use \xrade in the likelihood calculations for parameter inference in a Bayesian model. One solution is to integrate \xrade into the publicly available Bayesian software e.g., \textsc{3ml}\footnote{\url{https://threeml.readthedocs.io/en/stable/xspec_users.html}}. The advantage of such an approach is that we will be able to obtain more accurate parameter estimates and their full posterior distributions, all the while taking into account any principled prior information about the source.

The great advantages of \xrade are the following: (a) it avoids the need to rerun the initial time consuming ray tracing simulations, speeding up, in turn, the process of generating new spectra or even grids; (b) our \ann allows the user to generate fully customised \xrade tables at the user's specific requirements; (c) it produces very large \xrade tables, e.g., with much finer steps, over a much shorter computational timescale, i.e., seconds--minutes; (d) it greatly mitigates interpolation issues within \xspec between coarse grid points, while maintaining numerical accuracy to the 1\% level (see \autoref{fig:pcygni_emu}); and finally, (e) the emulation process can be applied to a large variety of models (see text below) and can be easily implemented directly into Bayesian inference pipelines. 

We presented a test case by applying \xrade and \dwf to \pds, which hosts one of the most powerful, persistent accretion disk-winds. We specifically tested \xrade on the combined \xmmnu 2013 September 17--21 observations of \pds, as the X-ray spectrum is characterized by the best-quality P-Cygni feature observed to date, and compared the results with those from \dwf. We found that both \xrade and \dwf return an excellent fit to the data, providing measurements of $\mw$, $\lx$, $\fv$, $\mu$ and $\Gamma$ with $<$\,10\% discrepancy. We demonstrated that \xrade provides an excellent fit to the P-Cygni profile in \pds. 

The best-fit values measured with both \dwf and \xrade are loosely consistent with N15; in particular the $\mw$ is a factor of $\sim3$ smaller than in N15. This difference can be simply attributed to an assumed launching radius being a factor of $\sim3$ larger i.e., $\rmin=100\rg=1.5\times10^{16}\,\rm cm$ than here. It is important to note that since $\rmin$ is not yet a free parameter but fixed a priori, the `true' mass outflow rate maintains a certain degree of uncertainty. For this reason, in future work it is our priority to make $\rmin$ a measurable parameter in \xrade, as well as to further explore the wind thickness ($\rmax/\rmin$) or even a variable $d$ parameter (i.e., changes the wind opening angle). Note that another source of discrepancy for $\mw$ can be also attributed, on a lower extent, to the assumed accretion efficiency value of $\eta=0.06$ here w.r.t. that in N15 (i.e., $\eta=0.1$).

The extended energy band from $0.1\kev$ up to $511\kev$ was adopted as in S10 in order to allow a comparison with observational measurements from future instruments with a significant effective area at relatively high photon energies, $>100\kev$. However, as such a milestone has not been achieved yet, a possibility for the near future would be to restrict the energy range of the calorimeter-resolution grids, so to optimize computational time and parameter space sampling over the region where this is most relevant (especially for covering the \fe region).

At present, the major difference between \xrade and disk-wind tables (\dws and \dwf) generated through a `standard' X-ray tracing method is the enormous difference of CPU time involved in the process. To emulate one single spectrum we require a CPU time of $4.9\times10^{-5}$ seconds, against $10$--$50$ minutes ($\sim$\,60 eV resolution) or 2--3 hours for ($2\ev$ resolution). We also used \bxa to perform a global exploration of the parameter spaces associated with the original \dwf and finer \xrade tables whilst fitting \pds (\autoref{subsec:bxa}). We find good agreement between the overall best-fitting parameter contours, as well as individual posterior distribution shapes (see \autoref{fig:bxa_pds456_corner}), indicating that the \ann is able to learn the complex mapping between parameter space and spectral space. Global parameter exploration algorithms thus represent a powerful tool to iteratively test the accuracy of emulation-based table models in the future.

Although \xrade is already a powerful alternative model to the computationally expensive \mcrt simulations, there is still much room for improvement. Most notably, the increase in fractional error seen in the \fe band will be improved by introducing finer sampling in the training process. Currently our training set is based on simulated spectra generated from a grid of parameters, however ideally we would train from spectra that have parameter values that are randomly sampled across the chosen parameter range. Using a random parameters allows the network to better map the domain and parameter space in comparison to the grid of parameters. Analogous to this, is the extensive research that has shown that random search is superior over grid search methods for hyper parameter turning of machine learning algorithms \citep[see e.g.][]{bergstra2012random}.
Any future work must allow for a sampling of $\mu$ and, most importantly, $\rmin$ values, so that a more accurate energetics and eventually the launching/driving mechanism involved in the disk-wind can be can be achieved. The real power of the emulation method is that the implementation of our \ann will undoubtedly be an indispensable tool in anticipation of future X-ray detectors, such as the micro-calorimeters on board \xrism and \athena. Our emulation method will not be only restricted to the development of \xrade, but it will be implemented in other wind models, such as magneto-hydrodynamic \citep[e.g.,][]{Fukumura10} and WINd Emission (WINE) models \citep{Luminari20}. This tool can be also applied to non-wind models and beyond X-ray astronomy studies.

\section{acknowledgements}
We would like to thank the anonymous referee for his/her helpful comments and feedback that helped us to improve the clarity of this paper. The authors would also like to thank Justin Alsing for helpful discussions. GAM acknowledges the financial support from Attività di Studio per la comunità scientifica di Astrofisica delle Alte Energie e Fisica Astroparticellare: Accordo Attuativo ASI-INAF n. 2017-14-H.0. GAM also acknowledges the Sciops technical IT Unit - SITU at the European Space Astronomy Centre for letting me use their cluster. ML acknowledges a Machine Learning in Science research fellowship from the University of Nottingham. JNR and VB acknowledges NASA-ADAP grant 80NSSC22K0474. PGB acknowledges financial support from the Czech Science Foundation project No. 22-22643S. ESK acknowledges financial support from the Centre National d’Etudes Spatiales (CNES). MB is supported by the European  Innovative Training  Network  (ITN)  ``BiD4BEST'' funded  by  the  Marie  Sklodowska-Curie Actions in Horizon 2020 (GA 860744). MG is supported by the ``Programa de Atracci\'on de Talento'' of the Comunidad de Madrid, grant number 2018-T1/TIC-11733. 

\section{Data Availability}
\dwf and \dws will be publicly available on \url{https://gabrielematzeu.com/disk-wind/}. \xrade models will be initially available on request to the authors and the \xrade generator will be publicly available in the foreseeable future. All \xmm and \nustar data used in this work are publicly
available from the corresponding archives.

\bibliographystyle{mn2e}
\bibliography{matzeu_references.bib}

\appendix


\section{Spectral properties of the disk-wind}
\label{secapp:Spectral properties of the disk-wind}

\subsection{Influence of geometry on the disk-wind features}
\label{subapp:Influence of geometry in disk-wind features}



In the \mcrt code, the photon packets are collected into $20$ inclination bins and then processed into 1000 energy bins. The observer's \los inclination ($\theta$) is measured with respect to the polar $z$--axis. Each angular bin is defined by $\mu=\cos\theta$ and determines the degree of \los interception through the wind. In both tables, the angular bins cover the range $0.025 <\mu < 0.975 $ in $20$ incremental linear steps of $\Delta\mu = 0.05$. 
As the geometric framework assumes a flow with an opening angle of $45^{\circ}$, the observer's \los does not directly intercept the wind when $\theta<45^{\circ}$, or $\mu\gtrsim 0.7$. In such a scenario, the corresponding spectra will be dominated by a reflection component via photons scattered off the wind (see \citealt{Tatum12a} for examples of fitting the wind spectra to the \iron emission profiles of bare Seyferts). Conversely, at high inclinations ($\mu\lesssim 0.7$), the \los intercepts the wind and, consequently, blueshifted absorption features, as well as scattered emission, will be imprinted on the spectra.

\begin{figure}
\centering
\includegraphics[scale=0.28]{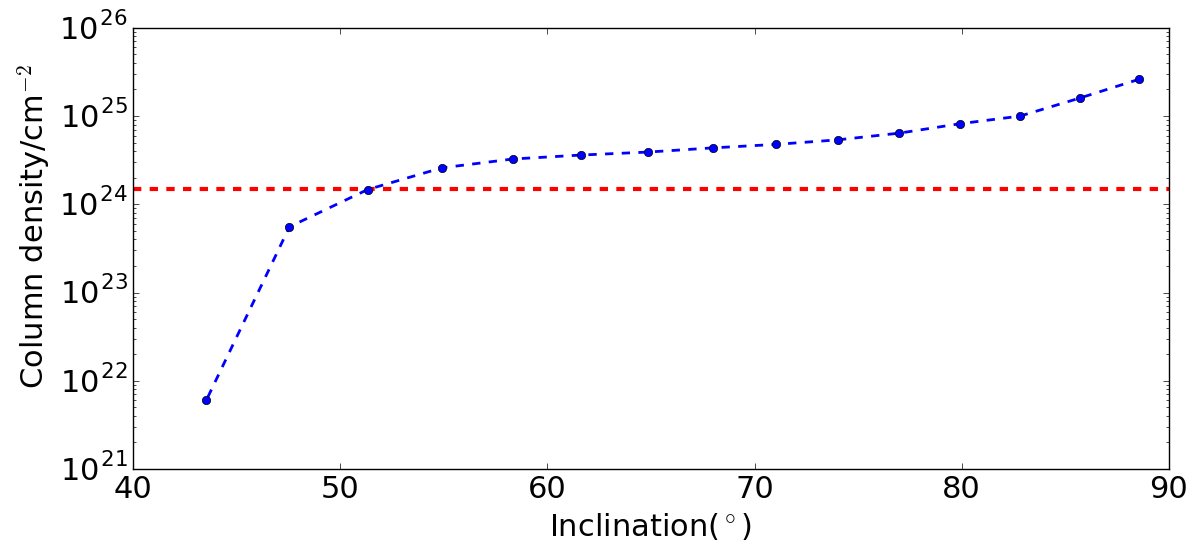}
\caption{Dependence of the \los column density on the viewing angle bins for a given $\mw=0.4$ in \dwf. Along a \los with $\theta<45^{\circ}$ the wind does not intercept the line of sight to the X-ray source so the column density is approaching zero, although the observed spectrum is modified by photons scattered into the line of sight. The column density increases for $\theta\gtrsim45^{\circ}$ as the line of sight becomes more edge-on. The dashed red line represents one Compton depth, corresponding to $\nh = 1/\sigma_{\rm T} = 1.5\times10^{24}$\,cm$^{-2}$, where the flux is suppressed to $38\%$ of the unattenuated flux.}
\label{fig:nhmu}
\end{figure}

Depending on the range of the angular bin, the inclinations can be denoted as: \textit{low} (polar; $\theta=0$--$45^{\circ}$), \textit{intermediate} (wind fully intercepted; $\theta=45$--$66^{\circ}$), and $high$ (edge-on or equatorial; $\theta=66$--$90^{\circ}$). The different sight lines, from each angular bin, intercept material with increasing column densities (or optical depth). \autoref{fig:nhmu} illustrates how the column density of the obscuring medium (for a given $\mw=0.4$ i.e., $30\%$ of $\medd$ in \dwf), rapidly reaches the optically-thick regime (i.e., $\nh=1/\sigma_{\rm T}=1.5\times10^{24}\cmsq$) with increasing $\theta$. The boundary at the opening angle $\theta=45^{\circ}$ would unavoidably create some discontinuity regions in the simulations. 

\begin{figure*}
\centering
\includegraphics[width=16cm]{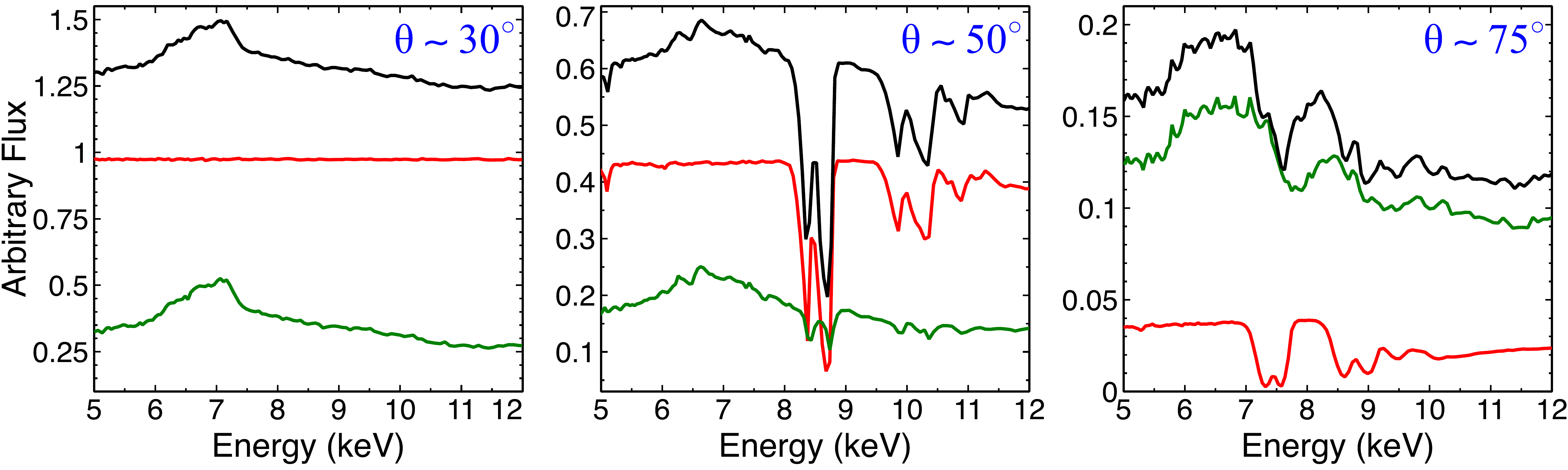}
\caption{Simulated \dwf spectra between $5$--$12\kev$ with given $\mw=0.5$, $\lx\,(\%)=0.2$ and $\fv=1.0$ corresponding to low (left; $\mu=0.875$, or $\theta\sim30^{\circ}$), intermediate (centre; $\mu=0.625$, or $\theta\sim50^{\circ}$) and high-inclinations (right; $\mu=0.275$, or $\theta\sim75^{\circ}$). The total spectra (black) and their respective direct (red) and scattered (green) components are shown. Low inclinations: as the \los does not intercept the wind, the resulting spectrum is dominated by scattered photons from the inner edge of the flow, leading to the broad \fe emission feature peaking at $E\sim7\kev$. Intermediate inclinations: as the \los fully intercepts the wind, deep absorptions e.g, from \fexxvxxvi, as well as broad emissions are imprinted on the spectra. High inclinations: at nearly equatorial \los orientation, the total spectrum is dominated by the scattering component whilst the primary continuum is heavily suppressed by the Compton-thick material at the base of the flow. Such orientation would lead to a broad, shallow absorption feature in the spectra.}
\label{fig:mu_zoomin}
\end{figure*}

In \autoref{fig:mu_zoomin} we show the output spectra from the different inclination ranges indicated above, where the total, direct (or transmitted), and scattered/reflected spectral components are denoted in black, red, and green, respectively. Specifically, at low-inclination e.g., $\theta\sim30^{\circ}$ (left), the \los does not intercept the wind, so the transmitted spectrum (red) is unaffected by the medium. The total spectrum (black) is dominated by the primary continuum and is supplemented by the scattered/reflected component (green) from the wind material. Distinct features such as the `Compton hump' (peaking at $20$--$30\kev$ ) and \feka emission at $\sim$\,6.4 keV (blurred by the Doppler shifts within the flow) are prominent. At intermediate values, $\theta\sim50^{\circ}$ (middle), the \los intercepts the outflowing Compton-thick material with $\nh\sim2\times10^{24}\cmsq$. The scattered component is similar to before, however the direct continuum is now suppressed by the obscuring medium with absorption features imprinted on the spectra. At high inclinations, $\theta\sim75^{\circ}$ (right), the scattered emission dominates over the transmitted component as the \los is intercepting material with $\nh\sim5\times10^{24}\cmsq$ (see \autoref{fig:nhmu}), corresponding to a Thomson (or Compton) depth of $\tau\sim3$. 

The difference in shape and centroid energy of the absorption profiles, as seen in \autoref{fig:mu_zoomin}\,(centre and right), reveals how the \fe strength and degree of blueshift are strongly dependent on both the wind opening angle and the \los orientation. In fact, at intermediate inclination, the centroid energy is measured at $\sim8.2\kev$ (centre), whilst at high inclination, the line is centred at $\sim7.2\kev$ (right). In other words, as the viewing angles progressively become polar the shift in velocity of the profile increases. Such variation arises from the \los projection of the velocity vector (see Fig.\,6, S08).

\subsection{Influence of the launch radius on wind features}
\label{sub:Launch Radius}

\begin{figure*}
  \includegraphics[width=16cm]{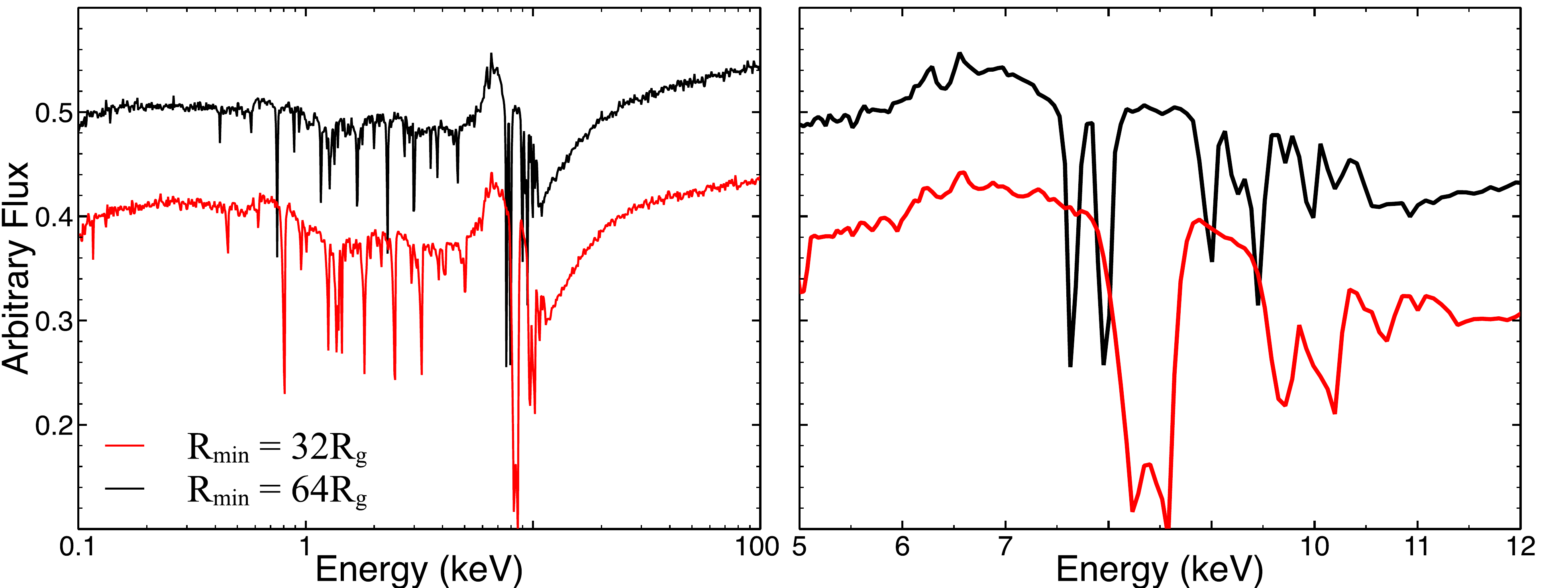}
  \caption{Left: Total broadband simulated spectra shown in the $0.1$--$100\kev$ band, with a given $\mu=0.625$, corresponding to launch radii of  $\rmin=64\rg$\,(black) and $\rmin=32\rg$\,(red). Right: Zoom on the \fe region. Regulated by the assumptions on $\rmin$, a velocity shift is observed in the synthetic spectra, whereby the black one ($\rmin=64\rg$) is slower than the red one ($\rmin=32\rg$). This is a natural consequence of the velocity along the streamline, which scales directly with the escape velocity at the base of the flow. As the terminal velocity $\vinf=\fv\sqrt{2/\rmin(\rg)}\,c$ is affected by both $\rmin$ and $\fv$ (or an {\it effective} $\rmin \propto f_{\rm v}^{2}$), in our simulations we kept $f_{\rm v} = 1$ for clarity. Additionally, with a given mass outflow rate ($\mw=0.5$) and ionizing luminosity ($\lx=2\times10^{-3}$, i.e., $L_{2-10\kev}=0.2\%$ of $\ledd$), the fastest wind spectrum (red) is overall more attenuated and with stronger line depths than its counterpart. The innermost winds have a larger column density as a consequence of the observer's \los crossing a larger portion of the wind than the outer flows (see Appendix\,\ref{sub:Mass density}).}
\label{fig:change_rin_spec}
\end{figure*}

The overall shape of the \fe absorption profile in the simulated spectra changes upon the choice of the launch radius $\rmin$ and thickness of the flow $\Delta R$. In \autoref{fig:change_rin_spec} we show the broadband simulated spectra for the \dws and \dwf grids with fine-tuning velocity factor parameter fixed at $\fv=1$. We find that: 
\begin{itemize}
    \item[(i)] the effect of $\rmin$ on the terminal velocity, and thus the degree of blueshift of the profiles, is larger at smaller radii;
    
    \item[(ii)] the width of the absorption lines depends on the range of the velocities intercepted. Broader profiles are thus naturally reproduced for faster (inner) winds -- due to both the larger range in terminal velocity and the wider shear of velocities intercepted along the flow up to $\vinf$ (see Appendix\,\ref{sub:velocity_subsection});
    
    \item[(iii)] an inner, faster wind has a greater opacity as its $\Delta R$ (in $\rg$) is smaller, thus the density is higher for any given $\mw$.
\end{itemize}


\subsection{Calculation of Wind velocity}
\label{sub:velocity_subsection}

A prescription that stipulates the rotational velocity at every point in the wind, following the parameterization of \citet{Knigge95}, was included in the S10 code. It is assumed that the specific angular momentum is conserved by the outflowing `packets' of matter about the polar $z$--axis. At the base of the wind streamline, the angular momentum of the packets is assumed to be Keplerian for the radius at which the streamlines cross the $xy$--plane. Thus the rotational velocity is solely defined by choosing the wind geometry and black hole mass of the source. 

The outflow radial velocity, which points away from the focus point of the wind $d$ (see Fig.~\ref{fig:Diskwind_KWDSchema}), is:
\begin{equation}
  v_l = v_0 + (v_{\infty} - v_0) \left(1 - \frac{R_v}{R_v + l} \right)^{\beta},
  \label{dw:vlaw}
\end{equation}

\noindent where $l$ is the distance along the wind streamline, and $R_{v}$ is the velocity scale length (set to be equal to $\rmax$), which defines how far the packet of matter has travelled before reaching halfway of the terminal speed in the streamline. The $\beta$ exponent governs the acceleration rate and is usually set to 1 due to difficulties in constraining it with the X-ray data currently available. The initial velocity $v_{0}$ is set to 0, given that $v_{0} \ll v_{\infty}$. Variations in the $R_{v}$ and $\beta$ terms in \autoref{dw:vlaw} can result in a change of the width of the simulated wind feature profiles. Reducing $\beta$ would increase the red-wing of the absorption feature, as the material would take longer to accelerate to $v_{\infty}$. On the other hand, by reducing $R_{v}$, the packet of matter would travel a shorter distance along the streamline before $v_{\infty}$ is reached, hence reducing the red-wing. Additionally, the presence and strength of a red-wing characterizes the probability of observing slower packets of matter along a given streamline.


\subsection{Mass density}
\label{sub:Mass density}
The wind is assumed to be smooth and in a steady state, and it can be characterized by a mass-loss rate $\mout$ which corresponds to the total mass present within the flow. The local mass-loss rate per unit area as a function of $R$ is defined as $d \dot m/dA \propto R^{\kappa}$. In these simulations the mass-loss exponent is set to $\kappa=-1$ (default value) which falls within the range expected in a continuous large-scale radial outflow i.e. $-1.3 < \kappa < -1$ \citep{Behar09}. The integral of $d \dot m/dA$ has to equal the total mass-loss rate, such that $\mout = 4 \pi \int_{\rmin}^{\rmax} (d\dot m/dA) R d(R)$. Thus,

\begin{equation} 
  \frac{d \dot m}{dA} = \frac{\mout (\kappa+2)}{4\pi[\rmax^{\kappa+2}-\rmin^{\kappa+2}]} R^{\kappa}.
  \label{dw:msurf}
\end{equation}

\noindent A decrease of the mass-loss parameter has the effect of making the mass within the flow more centrally concentrated. The mass density for a given cell is $\rho = dm/dV$, the unit volume is $dV = v_{l} dt dA $, where $v_l dt$ is the distance travelled by a packet of matter at velocity $v_l$ along the streamline. By combining these terms, the mass per unit volume is:

\begin{equation}
  \rho =\frac{1}{v_l}\frac{d \dot m}{dA}.
  \label{dw:rho}
\end{equation}

\begin{figure}
\centering
\includegraphics[width=\linewidth]{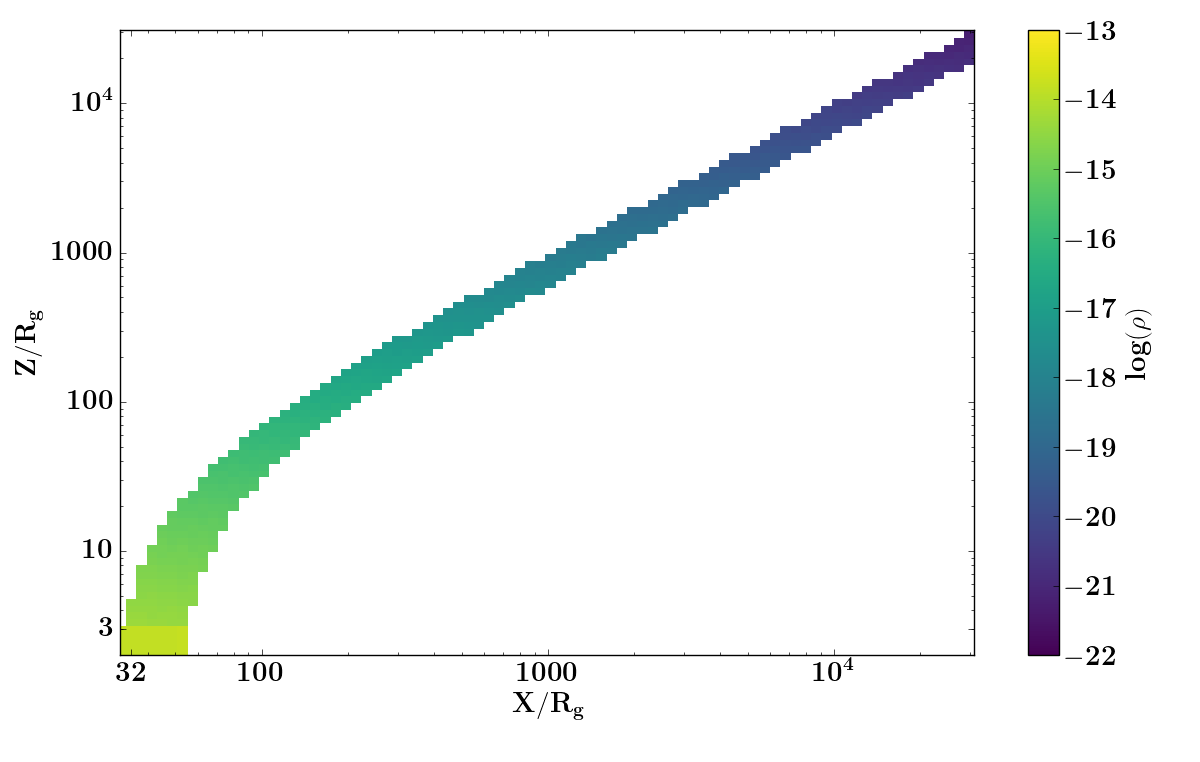}
\caption{Colour map of an example model run showing the distribution of the mass number density through the wind cells which falls off with radius as a function of $R^{(\kappa=-1)}$. The corresponding density values in log scale, are shown in the colour bar in units of g\,cm$^{-3}$. The $x$-axis represents the disk plane while the $z$-axis is along the rotational axis. Both axes are in units of gravitational radii.}
\label{fig:density}
\end{figure}

\noindent The above expression suggests that the mass density falls off faster than what is expected in \autoref{dw:msurf} at $\kappa = -1$, as the wind accelerates up to $v_{\infty}$. Such occurrence can be seen in \autoref{fig:density} where the mass number density at $R = 10^{3}\rg$ and $10^4\rg$ is $\log(\rho/\rm g\,\,cm^{-3})=-17.6$ and $\log(\rho/\rm g\,\,cm^{-3})=-20.1$ respectively i.e.,  $\Delta \log(\rho/\rm g\,\,cm^{-3}) = -2.5$. Thus the mass density falls off quicker than what would be expected from the mass loading \autoref{dw:msurf} at $\kappa = -1$ due the effect of the increasing velocity vector of the flow along the stream lines.

\section{Network parameters}
\label{sec:Network parameters}
 In a Feed-Forward Neural Network framework the number of trainable parameters ($N_p$) is derived from the number of connection in between each layer plus the number of biases in each layer. Each dense layer contains one bias per neuron, so we have a total of $3000$ biases. A general expression can be written as 

\begin{equation}
N_p=\sum_{k=1}^{n} N_{H^k} N_{H^{(k-1)}} + N_{H^k}
\end{equation}
where $n$ is the number of dense layers, $N_{H^0}$ is the number of inputs to the neural network. In our case $n=3$ and $N_{H^0}=6$. Each of of the 3 dense layers $H^1,H^2$ and $H^3$ have $1000$ neurons each so we have $N_p=(6\times1000 + 1000) + (1000\times1000 + 1000)+(1000\times1000 + 1000)=2,009,000$.

\end{document}